\documentclass{article}
\usepackage{graphicx}
\usepackage{float}
\usepackage{vmargin}
\usepackage{authblk}
\usepackage{amsmath}
\usepackage{amsfonts}
\usepackage{amssymb}
\usepackage{relsize}
\usepackage{bbm}
\usepackage{bm}
\usepackage{hyperref}
\usepackage{cleveref}
\usepackage{multirow}
\usepackage{lipsum}
\usepackage{cancel}

\usepackage{tikz}
\usetikzlibrary{arrows}
\usetikzlibrary{trees}
\usetikzlibrary{decorations.markings}
\usetikzlibrary{arrows.meta}

\tikzset{point/.style={insert path={ node[scale=6*sqrt(\pgflinewidth)]{.} }}}

\newcommand{\indicator}[2]{\mathlarger{\mathbbm{1}}\,_{#1}#2}

\def\independenT#1#2{\mathrel{\rlap{$#1#2$}\mkern2mu{#1#2}}}
\newcommand\independent{\protect\mathpalette{\protect\independenT}{\perp}}

\def\Y{\mathbf{Y}}

\def\V{\mathbb{V}}
\def\E{\mathbb{E}}

\def\R{\mathbb{R}}

\def\Normal{\mathcal{N}}

\newcommand{\llavs}[1]{\left\lbrace #1 \right\rbrace}       
\newcommand{\corchet}[1]{\left\lbrack #1 \right\rbrack}     
\newcommand{\parent}[1]{\left( #1 \right)}                  

\newcommand\keywords[1]{
    \renewcommand\thefootnote{}\footnote{#1}
    \addtocounter{footnote}{-1}
}

\title{Income, education, and other poverty-related variables: \\ a journey through Bayesian hierarchical models}
\author[1,*]{Irving Gómez-Méndez}
\author[2]{Chainarong Amornbunchornvej}
\affil[1,2]{National Electronics and Computer Technology Center (NECTEC)}
\affil[*]{Corresponding author: \href{gomendez.irving@gmail.com}{gomendez.irving@gmail.com}}
\affil[2]{\href{chainarong.amo@nectec.or.th}{chainarong.amo@nectec.or.th}}
\date{}

\begin{document}

\maketitle


\section*{Highlights}

\begin{itemize}
    \item One-shirt-size policy cannot handle poverty issues well since each region has its unique challenge while having custom-made policy for each region separately is unrealistic in term of resources.
    
    \item In this work, hierarchical models are deployed to explain income and poverty-related variables in the multi-resolution governing structural data of Thailand.
    
    \item Having a higher education level increases significantly the households' income for all the regions in Thailand.
    
    \item The impact of the region in the households' income is almost vanished when education level or years of education are considered. Therefore, education might have a mediation role between regions and income.

    \item For each year-of-education the monthly income of a household is increased in 1880 THB on average at the national level. This amount might vary between 1278 THB for Western Thailand up to 2545 THB for Eastern Thailand.

    \item Households with a high education level earn on average 8325 THB more per month than households with a mid education level and 14844 THB more than households with a low education level.
\end{itemize}

\begin{abstract}
    One-shirt-size policy cannot handle poverty issues well since each area has its unique challenges, while having a custom-made policy for each area separately is unrealistic due to limitation of resources as well as having issues of ignoring dependencies of characteristics between different areas. In this work, we propose to use Bayesian hierarchical models which can potentially explain the data regarding income and other poverty-related variables in the multi-resolution governing structural data of Thailand. We discuss the journey of how we design each model from simple to more complex ones, estimate their performance in terms of variable explanation and complexity, discuss models' drawbacks, as well as propose the solutions to fix issues in the lens of Bayesian hierarchical models in order to get insight from data. 
    
    We found that Bayesian hierarchical models performed better than both complete pooling (single policy) and no pooling models (custom-made policy). Additionally, by adding the year-of-education variable, the hierarchical model enriches its performance of variable explanation. We found that having a higher education level increases significantly the households' income for all the regions in Thailand.   The impact of the region in the households' income is almost vanished when education level or years of education are considered. Therefore, education might have a mediation role between regions and the income. Our work can serve as a guideline for other countries that require the Bayesian hierarchical approach to model their variables and get insight from data.
\end{abstract}

\section{Introduction}
Poverty is one of the most important issues that mankind faces~\cite{AMORNBUNCHORNVEJ2023e15947}. It is one of the main root causes that harms several aspects of society such as economy development~\cite{NAKABASHI2018445}, education~\cite{SILVALAYA2020100280}, healthcare systems~\cite{ibrahim2021health}, etc. For each year, there were millions of human deaths causing by poverty~\cite{pogge2005world}. To combat poverty issues, a government needs appropriate policies to solve them~\cite{zhang2023alleviating,okpala2023socio}. With the proper policies and sufficient resources, poverty can be alleviated effectively. However, finding the right policy is a non-trivial task due to the complexity of issues and unique characteristics of regions.  For instance, in a similar problem, one solution in a specific region might not be able to solve it in another region even though they have similar characteristics in many aspects~\cite{lahn2017curse,RIOJA2004429}.

One-shirt-size policy is a popular way to solve an issue by policy makers since it is simple to implement and typically uses less resources than a custom-made policy that is designed for a specific region. Nevertheless, one-shirt-size policy is unable to handle issues in all regions of a country effectively since each region might have their own unique socioeconomic context or other issues of poverty~\cite{berdegue2002rural,commins2004poverty,pringle2000cross}.  On the other hand, making a specific policy for each region to solve their unique problems is impossible due to the limitation of time and resources~\cite{10.1145/3424670}.

To find an optimal solution between the two extremes of one-shirt-size and custom-made policies, in this work, we propose to use \textit{Bayesian hierarchical models}~\cite{gelman2006data} to find a proper model that effectively explains target variables (e.g. income, debt, savings, etc.) related to poverty. To the best of our knowledge, there is no work in the literature that makes use of Bayesian hierarchical models to analyze variables of poverty in Thailand. For some variables, we found that complete pooling (representing one-shirt-size policy) and no pooling (representing custom-made policy) cannot explain a target variable while hierarchical models can, which represents the middle ground between these two extremes. The analyses and results of our work can be used as a role model for the analysis in other countries.

\subsection{Related works}
In the past, poverty was about lacking of income. However, the concept of poverty is complex and multidimensional~\cite{AMORNBUNCHORNVEJ2023e15947,wang2022relationship}, which implies that solving only monetary problems is not enough to alleviate the issues of poverty. Therefore, the  Multidimensional Poverty index (MPI)~\cite{alkire2021global,alkire2015multidimensional} was developed and used by United Nations Development Programme (UNDP) as a standard way to measure poverty in multidimensions such as living standard, health, education, etc. The MPI indices from  nations around the world have been reported annually by UNDP. Another index that is typically used for measuring income inequality is the Gini coefficient~\cite{asongu2023conditional}, which represents the distribution of resources among people. Instead of the Gini coefficient, the work in~\cite{amornbunchornvej2020nonparametric} proposed to use the network density of income gaps (edges represent significant gaps) to measure income inequality among different occupations. 

Even though these indices provide rich information regarding poverty and income inequality in each area, they never provide the information of resolution of poverty issues; given multiple areas, it is impossible to tell from MPI whether these areas share similar issues and need only a single policy to solve poverty. To address this gap, both minimum description length (MDL)~\cite{10.1145/3424670} and Gaussian Mixture Models~\cite{grun2007applications,grun2006fitting,JSSv011i08} can be used to find optimal multiresolution partitions that can place a single policy for each partition since each one represents an area that have a similar model of issues. However, these works cannot be used to provide insights regarding  dependencies of issues between different area resolution levels. Does income variables in the national level affects income variables in provinces or lower levels? The next section provides the reasons of using Bayesian hierarchical models in our work.

\subsection{Relevance of Bayesian hierarchical models}
One of the approaches to model policies is to use Bayesian's statistics and modeling, which is widely used to model public policies in government setting~\cite{fienberg2011bayesian} as well as public opinions~\cite{caughey_warshaw_2015}. In this work, we propose to use Bayesian hierarchical models to analyze variables that are related to poverty and inequality from a population dataset of Thai households.

Some datasets are collected with an inherent multilevel structure, for example, households within a region of a country. Then, hierarchical modeling is a direct way to include clusters at all levels of a phenomenon, without being overwhelmed with the problems of overfitting. At a practical level, hierarchical models are flexible tools combining partial pooling of inferences. They have been successfully involved in various practical problems, including biomedicine \cite{smith2003assessing,zhang2014spatio}, genetics \cite{vallejos2015basics,wang2019data}, ecology \cite{boehm2015spatial,royle2008hierarchical}, psychology \cite{lee2011cognitive}, among others. We refer to \cite{van2021bayesian} for a review on Bayesian modeling and a further list of their applications, including Bayesian hierarchical models.

The traditional alternatives to hierarchical modeling are complete pooling, in which differences between groups are ignored, and no pooling, in which data from different sources are analyzed separately. As we shall discuss, both these approaches have problems. However, the extreme alternatives can be useful as preliminary estimates.

The rest of the article is organized as follows. In \Cref{sec:Method} we explain our methodology to select the variables studied in this work, how simple and complex models interact between them, and the criterion used to compare different models. In \Cref{sec:HierRegion} we introduce the hierarchical model as a trade-off between no pooling and complete pooling models. In \Cref{sec:HierRegionAndEducation} we present a hierarchical model that incorporates two non-nested clusters. \Cref{sec:HierRegress} is devoted to Bayesian hierarchical regression. Finally, in \Cref{sec:Conclusions} we discuss the principal insights observed throughout this work.
\section{Methodology}
\label{sec:Method}

With the proliferation of Bayesian methods (see \cite{van2021bayesian} for a list of open Bayesian software programs), they have become easier to build and implement than to understand what they are doing. In an attempt to narrow this gap, in this work we present a comprehensive framework for hierarchical models. Thus, we do not only show how to implement these models, but also how to interpret the parameters according to the level where they belong in the hierarchy and their relation with other parameters.

Instead of starting directly with the hierarchical models, we begin with the extreme cases of no pooling and complete pooling. It is only after analyzing their implications and their lack to explain adequately certain aspects of the data, that we introduce the hierarchical models as a way to mitigate these problems. Thus, every time we introduce a new hierarchical model is always as an extension of a previous one.

Noninformative priors (also known as reference priors or objective priors) are notoriously difficult to derive for many hierarchical models. Thus, throughout this work, we present an approach in which simpler models are used for prior specification in more complex models. This contrasts with the most common approach to prior specification in which a prior distribution is selected because it has been previously used in the literature. Based on the assumption that the community of people using that prior are doing it for a good reason. However, as pointed out by \cite{simpson2017penalising}, most of these priors have been chosen for specific problems and might be inappropriate for others. Furthermore, as commented by \cite{mcelreath2018statistical}:
\begin{quote}
    There’s an illusion sometimes that default procedures are more objective than procedures that require user choice, such as choosing priors. If that’s true, then all ``objective'' means is that everyone does the same thing. It carries no guarantees of realism or accuracy.
\end{quote}

As commented by \cite{gelman2006}, hierarchical models allow a more ``objective'' approach to inference by estimating the parameters of prior distributions from data rather than requiring them to be specified using subjective information. Moreover, in hierarchical models where priors
depend on hyperparameter values that are data-driven avoids the direct problems linked to double-dipping \cite{van2021bayesian}. Therefore, our approach follows the tendency by part of the Bayesian community to move from noninformative priors \cite{gelman2006data,mcelreath2018statistical,lemoine2019moving,gelman2013bayesian}. We do not claim that the proposed approach is optimal. Instead, we make the more modest claim that it is useful for practical purposes.

\subsection{Data and related information}

The dataset used throughout this work has been collected in 2022 by Thai government agencies. The main purpose of this dataset is for supporting government policy makers to calculate MPI to support poverty alleviation policy making in the Thai People Map and Analytics Platform project (\href{https://www.tpmap.in.th/about_en}{www.TPMAP.in.th})~\cite{10.1145/3424670}. The original dataset has 12,983,145 observations (each one corresponding to a household). However, there were 569 households that declared having no-income, which represents 0.0044\% of the observations. We consider this percentage negligible and  do not consider them for further analyses. The number of households consulted for each province goes from 44,012 up to 645,433, then, for all practical purposes, we can ignore the uncertainty within each province.

\subsection{Comparison of models}

Since different models are implemented for the same data and variables, we need to develop a methodology that allows us to compare them. For this purpose, we considered the Widely Applicable Information Criterion (WAIC), renamed in \cite{watanabe2010asymptotic} as the Watanabe-Akaike Information Criterion. This information criterion allows a fair comparison between models of different complexity. Compared to other information criteria like the Akaike Information Criterion (AIC) \cite{akaike1973information}, WAIC averages over the posterior distribution rather than conditioning on a point estimate (like the maximum likelihood estimator), making it a more suitable criterion for Bayesian models. Moreover, AIC is defined relative to the maximum likelihood estimate and so is inappropriate for hierarchical models. In \Cref{subsec:WAIC} we provide further details about the WAIC.

We emphasize that the WAIC is just another statistical summary of our models, and it is not, in any way, a substitute of an appropriate analysis of the models and their results. If we find that the model does not fit for its intended purposes, we are obliged to search for a new model that fits. Then, understanding different aspects of the models and their implications must be the principal guide for their selection and comparison. See \cite{aki2012} for further discussion on Bayesian predictive model assessment, selection, and comparison methods.

\subsection{Variables to analyze}
\label{subsec:SelectVar}

\begin{figure}[ht]
	\centering
	\includegraphics[width=0.65\textwidth]{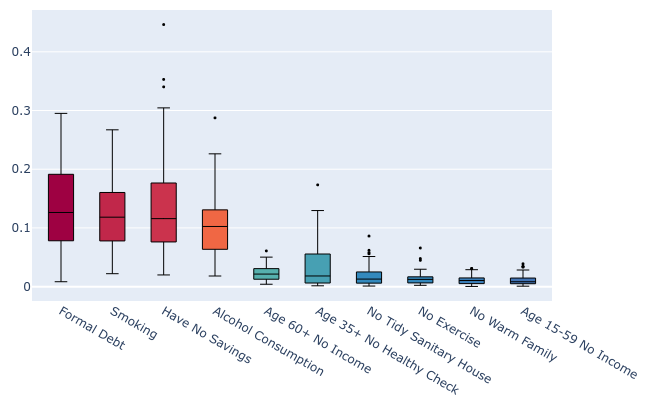}
	\caption{Boxplots of the 10 poverty-related variables affecting the largest percentage of households in Thailand.}
	\label{fig:Boxplots}
\end{figure}

For its direct relation with poverty, we exemplify our methodology throughout this work considering the monthly average income in households. However, this approach is suitable to be applied to a widely variety of variables. To decide the variables where to apply our methodology, we first calculated the percentage of households in each province affected by the variables considered in the dataset. In \Cref{fig:Boxplots}, we present boxplots for the 10 issues with the largest percentage.

Due to the large percentage of households affected by (having) formal debt, smoking, having no savings, and alcohol consumption, as well as their large variance between provinces, we also analyzed these variables with the different models presented in this work. The WAIC for the models implemented on these variables are presented in \Cref{tab:WaicAllModels}.
\section{Hierarchical model with one cluster: income per region}
\label{sec:HierRegion}

\subsection{No pooling model and complete pooling model}
\label{subsec:SeparateAndPooled}

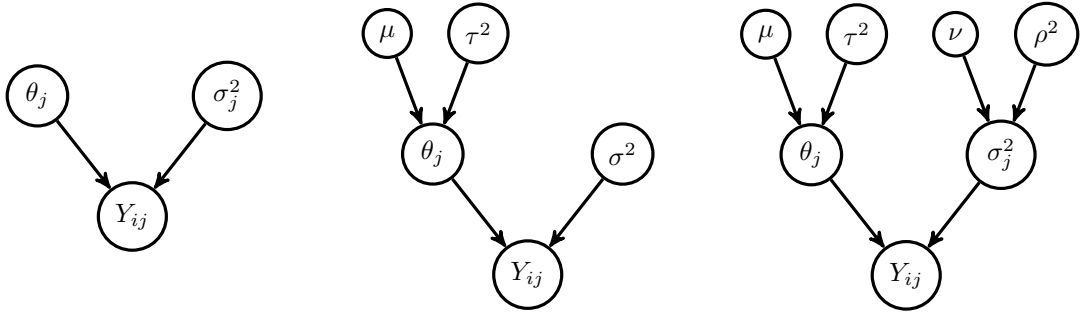
\begin{figure}[ht]
    \centering
	    \begin{minipage}[c]{0.32\textwidth}
            \centering
    	    \begin{tikzpicture}[>=stealth', grow=up,shape=circle,very thick,level distance=16mm,every node/.style={draw=black,circle},edge from parent/.style={<-,draw}]
                
                \tikzstyle{level 1}=[sibling distance=25mm]
    
        	    \node {$Y_{ij}$}
        		      child {node {$\sigma_j^2$}}
        		      child {node {$\theta_j$}};
                
    	    \end{tikzpicture}
        \end{minipage}
        \begin{minipage}[c]{0.32\textwidth}
            \centering
            \begin{tikzpicture}[>=stealth', grow=up,shape=circle,very thick,level distance=16mm,every node/.style={draw=black,circle},edge from parent/.style={<-,draw}]
            
    	        \tikzstyle{level 1}=[sibling distance=25mm]
                \tikzstyle{level 2}=[sibling distance=12mm]
    
            	\node {$Y_{ij}$}
            		child {node {$\sigma^2$}}
            		child {node {$\theta_j$}
            			child {node {$\tau^2$}}
            			child {node {$\mu$}}
            		};
              
    	    \end{tikzpicture}
        \end{minipage}\hfill
        \begin{minipage}[c]{0.32\textwidth}
            \centering
            \begin{tikzpicture}[>=stealth', grow=up,shape=circle,very thick,level distance=16mm,every node/.style={draw=black,circle},edge from parent/.style={<-,draw}]
            
            	\tikzstyle{level 1}=[sibling distance=25mm]
            	\tikzstyle{level 2}=[sibling distance=12mm]
            
            	\node {$Y_{ij}$}
            		child {node {$\sigma_j^2$}
            			child {node {$\rho^2$}}
            			child {node {$\nu$}}
            		}
            		child {node {$\theta_j$}
            			child {node {$\tau^2$}}
            			child {node {$\mu$}}
            		};
              
        	\end{tikzpicture}
        \end{minipage}
    \caption{Graphical representation of the considered models. Left: Separate independent models for each region, if we add the restrictions $\theta_j=\theta$ and $\sigma^2_j=\sigma^2$ for all $j$, then we get the complete pooling model. Center: Hierarchical model, where the regional means share a common structure, yet allowing them to be different, but the constraint of equal variance is still present. Right: Hierarchical model imposing a common structure for both, the regional means and the within-region variances.}
    \label{fig:GraphicalRepresentationModels}
\end{figure}

\subsubsection{No pooling model}

Let $Y_{ij}$ be the average household in the province $i$, which belongs to the region $j$, each region with $n_j$ observations. We consider 6 regions for Thailand: Northern Thailand, Northeast Thailand, East Thailand, Central Thailand, Western Thailand and Southern Thailand.

Before jumping directly into our hierarchical model, we first consider separate independent models for each region. Thus, each region has its own mean $\theta_j$ and its own variance $\sigma^2_j$. We assume that $Y_{ij}|\theta_j,\sigma_j^2\sim\Normal(\theta_j,\sigma_j^2)$\footnote{$\Normal(\theta,\sigma^2)$ denotes a normal distribution with location $\theta$ and scale $\sigma$.}, $j=1,\ldots,J$. This model is represented graphically on the left of \Cref{fig:GraphicalRepresentationModels}. For a simple model like this, we can consider vague noninformative priors without any harm, thus we use the well-known noninformative prior\footnote{$\indicator{A}{(x)}$ denotes the indicator function, defined as
\[
\indicator{A}{(x)}=
\begin{cases}
    1 &\text{ if } x\in A \\
    0 &\text{ if } x\notin A
\end{cases}
\]
for some set $A$ where $x$ is properly defined.} \[p(\bm{\theta},\bm{\sigma}^2)\propto \prod_{j=1}^J\frac{1}{\sigma_j^2}\indicator{\R}{(\theta_j)}\indicator{(0,\infty)}{(\sigma_j^2)}.\]

It is not difficult to prove that the conditional posterior distributions for each $\theta_j$ and $\sigma_j^2$ are given by\footnote{Inverse-$\chi^2(\nu,\sigma^2)$ denotes a scaled inverse $\chi^2$ distribution with $\nu$ degrees of freedom and scale $\sigma^2$.}
\begin{align*}
    \theta_j|\bm{\sigma}^2,\Y & \sim\Normal(\bar{Y}_{\cdot j},\bar\sigma_j^2), \\
    \sigma_j^2|\bm{\theta},\Y & \sim\textsf{Inverse-}\chi^2(n_j,v_j),
\end{align*}
where \[\bar{Y}_{\cdot j} = \frac{1}{n_j}\sum_{i=1}^{n_j}Y_{ij},\quad \bar\sigma_j^2=\frac{\sigma_j^2}{n_j},\quad \text{and }v_j=\frac{1}{n_j}\sum_{i=1}^{n_j}(Y_{ij}-\theta_j)^2.\]
Which makes it straightforward to simulate from the joint posterior distribution using Gibbs sampling \cite{geman1984stochastic}.

\begin{figure}[ht]
	\centering
	\begin{minipage}[c]{0.49\textwidth}
		\centering
		\includegraphics[width=\textwidth]{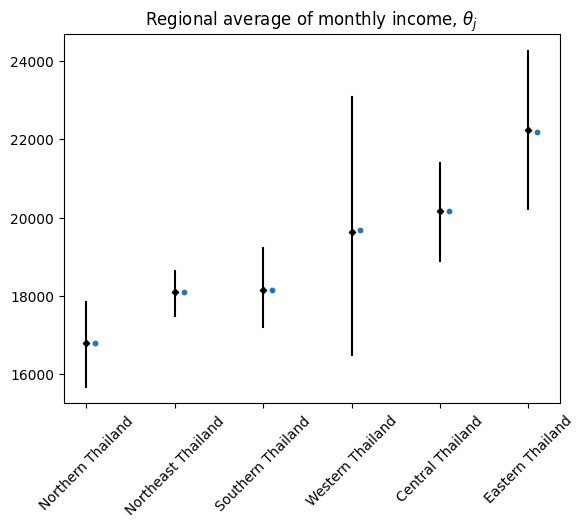}
	\end{minipage}\hfill
	\begin{minipage}[c]{0.49\textwidth}
		\centering
		\includegraphics[width=\textwidth]{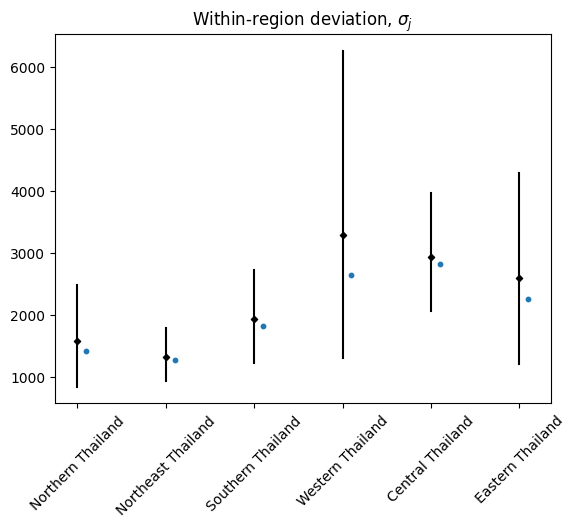}
	\end{minipage}
    \begin{minipage}[c]{0.49\textwidth}
		\centering
		\includegraphics[width=\textwidth]{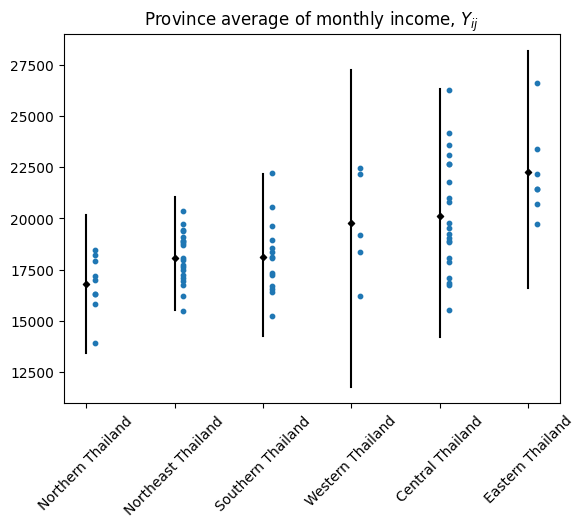}
	\end{minipage}
    \caption{Results considering independent separate models for each region. On the top row we show the regional mean, $\theta_j$ (left) and the regional standard deviation, $\sigma_j$ (right). On the bottom we show the province mean.}
    \label{fig:ResultsSeparateModels}
\end{figure}

On the top left of \Cref{fig:ResultsSeparateModels}, we present the estimated mean for each region, $\theta_j$, with a credible interval of 0.95 posterior probability. We observe that, since we considered noninformative priors, the estimations are centered on the observed regional averages. Note also that, because the regions share no information between them, credible intervals are large, especially for regions with few provinces. On the top right, we present a similar plot for the standard deviation for each region, $\sigma_j$. On the bottom, we present credible intervals for the average monthly income in each region. We present these credible intervals with the observed average for the provinces belonging to the region.

\subsubsection{Complete pooling model}

We can observe in \Cref{fig:ResultsSeparateModels} that the intervals overlap for most of the regions. This overlapping suggests that all the parameters might be estimating the same quantity. In fact, it is highly unlikely that the regions are independent between them, which makes difficult to justify an independent model for each one. Thus, we can consider the complete pooling model, in which all the regional means, $\theta_j$, and their variances, $\sigma^2_j$, are equal to some common values $\theta$ and $\sigma^2$, respectively. That is, for the complete pooling model, we assume that $Y_{ij}|\theta,\sigma^2\sim\Normal(\theta,\sigma^2)$. Similar to the no pooling model, the complete pooling is represented by the graph on the left of \Cref{fig:GraphicalRepresentationModels}, with the constraints that $\theta_j=\theta$ and $\sigma_j^2=\sigma^2$ for all $j=1,\ldots,J$.

For this model we still consider the noninformative prior distribution for $\theta$ and $\sigma^2$,
\[p(\theta,\sigma^2)\propto \frac{1}{\sigma^2}\indicator{\R}{(\theta)}\indicator{(0,\infty)}{(\sigma^2)}.\] It is not difficult to prove that the conditional posterior distribution of $\theta$ is given by $\theta|\sigma^2,\Y\sim\Normal(\bar{Y}_{\cdot \cdot},\varphi^2)$, where 
\[\bar{Y}_{\cdot\cdot} = \frac{\sum_{j=1}^J \frac{\bar{Y}_{\cdot j}}{\bar\sigma_j^2}}{\sum_{j=1}^J \frac{1}{\bar\sigma_j^2}},\quad \varphi^2 = \frac{1}{\sum_{j=1}^J \frac{1}{\bar\sigma_j^2}},\quad\text{and }\bar\sigma_j^2=\frac{\sigma^2}{n_j},\]
while the conditional posterior distribution of $\sigma^2$ is given by $\sigma^2|\theta,\Y\sim\textsf{Inverse-}\chi^2(n,\hat{\sigma}^2)$, with \[n=\sum_{j=1}^J n_j,\quad\text{and } \hat{\sigma}^2=\frac{1}{n}\sum_{j=1}^J\sum_{i=1}^{n_j}(Y_{ij}-\theta)^2.\]
Once again, having access to the conditional posterior distributions allows us to simulate from the joint posterior distribution using Gibbs sampling.

\begin{figure}[ht]
	\centering
	\begin{minipage}[c]{0.49\textwidth}
		\centering
		\includegraphics[width=\textwidth]{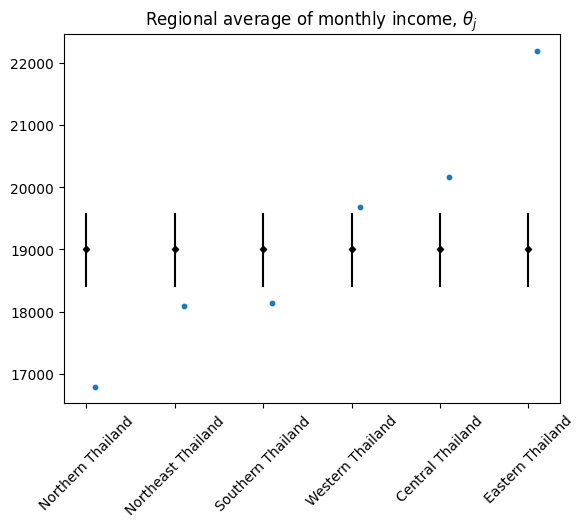}
	\end{minipage}\hfill
	\begin{minipage}[c]{0.49\textwidth}
		\centering
		\includegraphics[width=\textwidth]{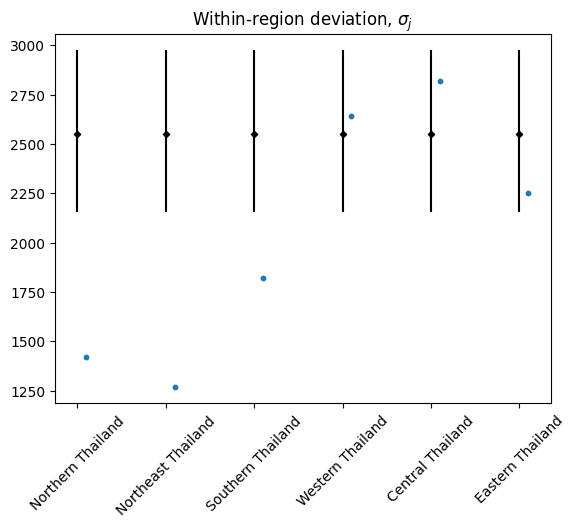}
	\end{minipage}
    \begin{minipage}[c]{0.49\textwidth}
		\centering
		\includegraphics[width=\textwidth]{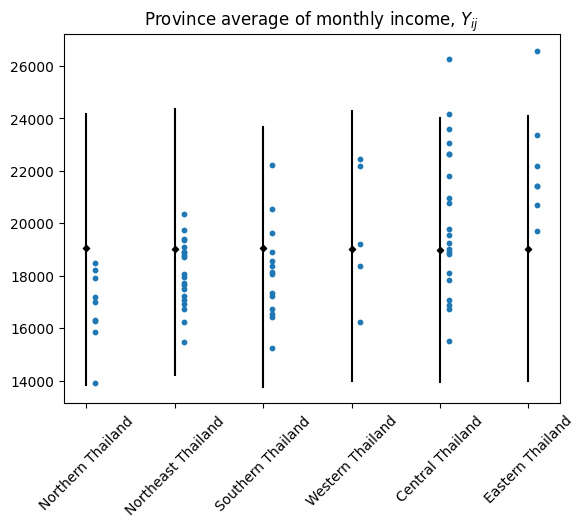}
	\end{minipage}
    \caption{Results considering the complete pooling model for all the regions. On the top row we show the regional mean, $\theta_j$ (left) and the regional standard deviation, $\sigma_j$ (right). On the bottom we show the province mean.}
    \label{fig:ResultsPooledModel}
\end{figure}

We present in \Cref{fig:ResultsPooledModel} the analogous results of \Cref{fig:ResultsSeparateModels} for the complete pooling model. Comparing both Figures, we observe much narrower intervals (also calculated at a 0.95 posterior probability), this is because now we are using all the observations to estimate the same common quantities, reducing the uncertainty significantly. However, we observe that the common mean $\theta$ can barely explain the mean of a few regions, being an unreliable estimate for the regions with the largest and smallest means. We can also observe that the estimator of the common within-region deviation is not centered around the average of the observed sample deviations, but upward. This is because now that we have constraint the regional means to be all the same, the only way to explain the variation throughout the observations is by estimating a higher value for the common deviation $\sigma$. Note also, on the bottom row of \Cref{fig:ResultsPooledModel}, that all the provinces in Northern Thailand are on the below half of the credible interval, while all the provinces in Eastern Thailand are above. We can conclude from all these observations that a complete pooling model is inappropriate.

\subsection{Hierarchical model with common within-cluster variance}
\label{subsec:HierCommonSigma}

Because considering independent models for each region seems difficult to justify and we observe a poor performance for the complete pooling model, we consider as a better approach a model that makes a trade-off between these two extreme cases. A hierarchical model achieves this compromise.

Instead of adding a hierarchical structure to all the parameters, we propose to add it to one parameter first, and consider more complex models only as suggested by the data after analyzing the results of the previous model. For this reason, we maintain a common within-region variance $\sigma^2$, but consider different regional means, $\theta_j$. However, these means are not independent, instead they share a common structure. From a statistical perspective this means to abandon the noninformative prior for $\theta_j$ and consider a distribution that depends on some hyperparameters, as it is represented by the graph in the center of \Cref{fig:GraphicalRepresentationModels}.

For the simplicity of a conjugate model \cite{gelman2013bayesian}, we consider the prior $\theta_j|\mu,\tau^2\sim\Normal(\mu,\tau^2),\quad j=1,\ldots,J$. In this model, $\mu$ represents the national average of the monthly income and $\tau$ represents the between-regions deviation. For the within-regions variance, $\sigma^2$ we still consider the noninformative prior \[p(\sigma^2)\propto\frac{1}{\sigma^2}\indicator{(0,\infty)}{(\sigma^2)}.\] 

To complete our model, we must assign prior distributions for $\mu$ and $\tau$. However, we must be careful since the usual noninformative distributions for location and scale parameters might lead to the non existence of the posterior distributions. For example, using the usual noninformative prior of a variance parameter for $\tau^2$, $p(\tau^2)\propto\frac{1}{\tau^2}\indicator{(0,\infty)}{(\tau^2)}$ yields an improper posterior distribution. Meanwhile, the vague prior \[p(\tau^2)\propto\indicator{(0,\infty)}{(\tau^2)}\] generates a proper posterior distribution, thus we use this prior for $\tau^2$. For $\mu$, we use the usual noninformative prior \[p(\mu)\propto\indicator{\R}{(\mu)}.\]

In \Cref{Subsec:AppQuali}, we present an empirical approach (developed in \cite{gelman2013bayesian}) to estimate these parameters, and explore in more detail the qualitative implications of the priors and the values taken by the parameters in the hierarchical models.

With this model, the following conditional distributions for the parameters can be deduced \cite{gelman2013bayesian}.

\paragraph{Conditional posterior for $\theta_j$}
\[\theta_j|\mu,\tau^2,\sigma^2,\Y\sim\Normal(\hat{\theta}_j,V_{\theta_j}),\]
where
\begin{equation}
\label{eq:HatThetaAndVTheta}
\hat{\theta}_j = \frac{\frac{1}{\bar\sigma_j^2}\bar{Y}_{\cdot j}+\frac{1}{\tau^2}\mu}{\frac{1}{\bar\sigma_j^2}+\frac{1}{\tau^2}},\quad\text{and } V_{\theta_j} = \frac{1}{\frac{1}{\bar\sigma_j^2}+\frac{1}{\tau^2}}.
\end{equation}

\paragraph{Conditional posterior for $\mu$}
\[\mu|\bm{\theta},\tau^2,\sigma^2,\Y\sim\Normal(\hat{\mu},\tau^2/J),\]
where $\hat{\mu}=\frac{1}{J}\sum_{j=1}^J\theta_j$.

\paragraph{Conditional posterior for $\sigma^2$}
\[\sigma^2|\bm{\theta},\mu,\tau^2,\Y\sim\textsf{Inverse-}\chi^2(n,\hat{\sigma}^2),\]
where $n=\sum_{j=1}^J n_j$, and \[\hat{\sigma}^2=\frac{1}{n}\sum_{j=1}^J\sum_{i=1}^{n_j}(Y_{ij}-\theta_j)^2.\]

\paragraph{Conditional posterior for $\tau^2$}
\[\tau^2|\bm{\theta},\mu,\sigma^2,\Y\sim\textsf{Inverse-}\chi^2(J-1,\hat{\tau}^2),\]
where \[\hat{\tau}^2=\frac{1}{J-1}\sum_{j=1}^J(\theta_j-\mu)^2.\]

\begin{figure}[ht]
	\centering
	\begin{minipage}[c]{0.49\textwidth}
		\centering
		\includegraphics[width=\textwidth]{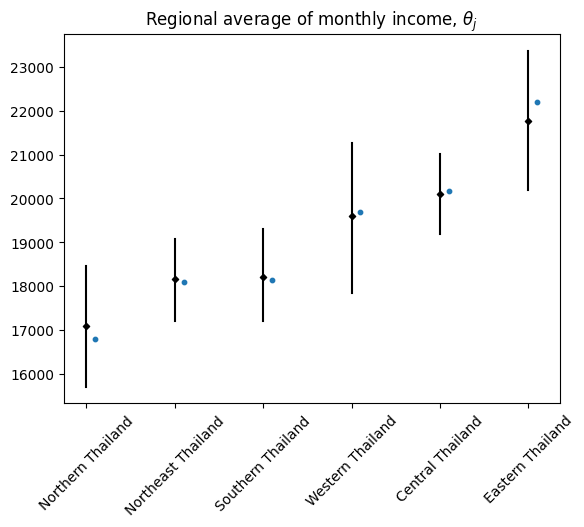}
	\end{minipage}\hfill
	\begin{minipage}[c]{0.49\textwidth}
		\centering
		\includegraphics[width=\textwidth]{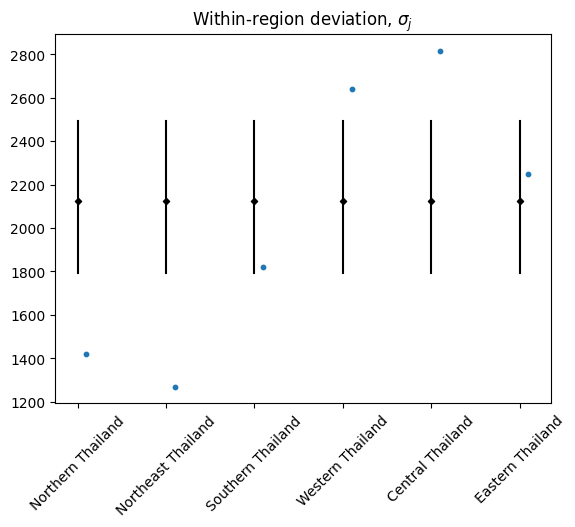}
	\end{minipage}
    \begin{minipage}[c]{0.49\textwidth}
		\centering
		\includegraphics[width=\textwidth]{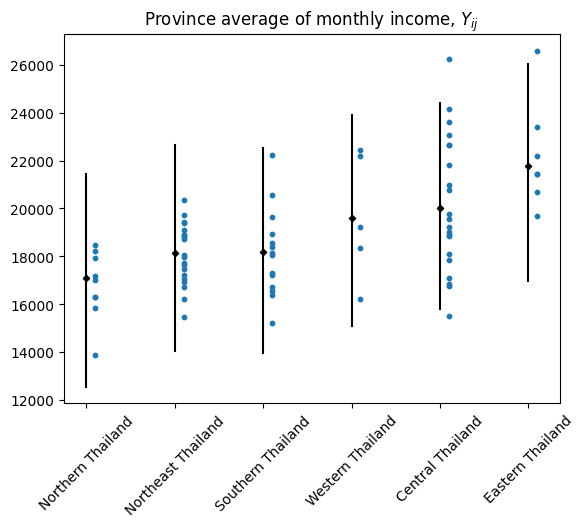}
	\end{minipage}
    \caption{Results considering a hierarchical model. We allow the mean of each region to vary while maintaining the same within-region variance $\sigma^2$. On the top row we show the regional mean, $\theta_j$ (left) and the regional standard deviation, $\sigma_j$ (right). On the bottom we show the province mean.}
    \label{fig:ResultsHierarchicalCommonSigma}
\end{figure}

As we did with the previous two models, we present in \Cref{fig:ResultsHierarchicalCommonSigma} the estimated mean for each region, $\theta_j$, the estimated common within-region deviation $\sigma$, and the average province monthly income, all of them with their respective credible intervals of 0.95 posterior probability. For the regional means, we can observe that the uncertainty is considerable less that when we considered independent analyses (see \Cref{fig:ResultsSeparateModels}) without having a poor performance as the complete pooling model. Also, since we accept different means for each region, now the common variance is not overestimated. However, we can observe that a common variance is infeasible to explain the observed variability for most of the regions.

\subsection{Hierarchical model varying within-cluster variance}
\label{subsec:HierVaryingSigma}

Because a common within-region variance $\sigma^2$ seems infeasible, we can impose a hierarchical level to it, similarly as we did with the regional means. Thus, we allow each region to have its own variance $\sigma_j^2$, but all of them sharing a common structure. For simplicity of a conjugate model, we consider the following prior distribution \[\sigma_j^2|\nu,\rho^2\sim\textsf{Inverse-}\chi^2(\nu,\rho^2).\]

This model is represented graphically on the right of \Cref{fig:GraphicalRepresentationModels}. These graphical representations, called Bayesian networks, meet two objectives. First, they visualize easily the hierarchical relations between variables, which helps with the interpretation of the parameters and the understanding of the model. Second, they allow us to use $d$-separation rules \cite{verma1990causal} to deduce the conditional independence between the parameters. We suggest \cite{barber2012bayesian} and \cite{pearl2000models} for gentle introductions to Bayesian networks and $d$-separation.

Consider, for example, the graph presented on the right of \Cref{fig:GraphicalRepresentationModels}, while \textit{a priori} $\theta_j$ is independent of $\sigma_j^2$, we can see that conditioning on $\Y$ creates a dependence between both parameters, that is $\theta_j\independent\sigma_j^2$, but $\theta_j\not\independent\sigma_j^2|\Y$. However, if we condition on both $\Y$ and $\sigma_j^2$, $\theta_j$ is independent of $\nu$ and $\rho^2$. This implies that the full conditional posterior of $\theta_j$ is exactly the same as in the model of \Cref{subsec:HierCommonSigma}, which assumes the same variance for all the regions (with the minor change of defining $\bar{\sigma}^2_j=\sigma^2_j/n_j$ instead of $\bar{\sigma}^2_j=\sigma^2/n_j$). Using the same reasoning, it is easy to see that the posterior distributions of $\mu$ and $\tau^2$ are also the same as those presented in \Cref{subsec:HierCommonSigma}.

Therefore, we only need to calculate the posterior distributions of $\sigma_j^2$, $\nu$ and $\rho^2$. Due to the conjugacy property of the model, it is not difficult to prove (see \cite{gelman2013bayesian}) that
\[\sigma_j^2|\bm{\theta},\nu,\rho^2,\Y\sim\textsf{Inverse-}\chi^2(\nu_j,\hat{\sigma}_j^2),\]
where
\[\nu_j=\nu+n_j,\quad \hat{\sigma}_j^2=\frac{\nu\rho^2+n_jv_j}{\nu+n_j},\text{and }v_j=\frac{1}{n_j}\sum_{i=1}^{n_j}(Y_{ij}-\theta_j)^2.\]

To complete our model, we must assign prior distributions for $\rho^2$ and $\nu$. Note that, considering the Bayesian network associated to the model, and $d$-separation, it is easy to see that once we condition on $\bm{\sigma^2}$, $\rho^2$ is independent of all the other variables and parameters of the model, except for $\nu$. For $\rho^2$, we can prove (see \Cref{subsubsec:PosteriorRho2}) that the vague prior \[p(\rho^2)\propto \frac{1}{\rho^2}\indicator{(0,\infty)}{(\rho^2)}\] yields the conditional posterior\footnote{Gamma$(\alpha,\beta)$ denotes a gamma distribution with shape parameter $\alpha$ and rate parameter $\beta$.} \[\rho^2|\bm{\sigma^2},\nu\sim\textsf{Gamma}\parent{\frac{J\nu}{2}, \frac{J\nu}{2\hat{\rho}^2}},\] where \[\hat{\rho}^2 = \frac{J}{\sum_{j=1}^J \frac{1}{\sigma_j^2}}.\]

Unfortunately, the conditional posterior of $\nu$ is far more complicated, let be $\omega=\nu/2$, then
\[p(\nu|\bm{\sigma}^2,\rho^2) \propto p(\nu)\frac{\omega^{J\omega}}{\Gamma^J(\omega)}(\rho^2)^{J\omega}\exp\llavs{-J\omega\frac{\rho^2}{\hat{\rho}^2}}\prod_{j=1}^J(\sigma_j^2)^{-\omega}.\]
This is an intricate expression, which gives little guide for the selection of a prior distribution for $\nu$ yielding a proper known distribution.

In general, it remains challenging to propose noninformative priors for the degrees of freedom of a distribution. In \cite{VILLA2018197} the authors present different proposals for the degrees of freedom of a $t$-distribution under certain conditions. Notably, in \cite{simpson2017penalising}, the authors present a process to build objective priors, which they called \textit{Penalised Complexity} or PC priors. Let as it be, these approaches do not necessarily applied for the degrees of freedom of an Inverse-$\chi^2$ distribution. Thus, in this work we consider three different approaches to propose a prior distribution for $\nu$.

\subsubsection{Estimating \texorpdfstring{$\rho^2$}{rho2} and \texorpdfstring{$\nu$}{nu}: the hierarchical model with fixed \texorpdfstring{$\hat\nu$}{nu}}

Because $\sigma_j^2|\nu,\rho^2\sim\textsf{Inverse-}\chi^2(\nu,\rho^2)$, we can use the method of moments to estimate $\rho^2$ and $\nu$. Let be $E_{s^2}$ the average of the observed sample within-group variances, $s_1^2,\ldots,s_J^2$, and $V_{s^2}$ their variance, using the method of moments (see \Cref{subsubsec:EstimateRhoNu}), we get the following estimates
\begin{equation}
  \hat{\nu}=\frac{2(E_{s^2})^2}{V_{s^2}}+4
  \label{eq:HatNu}
\end{equation}
and
\begin{equation}
    \hat{\rho}^2=\parent{\frac{2(E_{s^2})^2+2V_{s^2}}{2(E_{s^2})^2+4V_{s^2}}}E_{s^2}
    \label{eq:HatRho}
\end{equation}
Thus, the first option considered in this work is to fix the value of $\nu$ to its empirical estimator.

\subsubsection{Using a vague improper prior for \texorpdfstring{$\nu$}{nu}}
Setting $\nu$ to a fix value like $\hat{\nu}$ ensures us that the posterior distribution would exist for all the parameters, except $\nu$ which is no longer modeled as a random variable. This means that, setting the value of $\nu$ to a fix value, eliminates the uncertainty that we have on that parameter, and makes our model overconfident because it acts as if $\hat\nu$ would be the real value of $\nu$. For this reason, the second approach in this work is to consider a vague prior for $\nu$.

For modeling the degrees of freedom of a multivariate $t$-distribution, \cite{anscombe1967topics} proposed the prior $p(\nu)\propto(\nu+1)^{}-3/2$, while \cite{relles1977statisticians} proposed to use $p(\nu)\propto\nu^{-2}$. Thus, we proposed a prior for $\nu$ of the form $p(\nu)\propto \nu^{-h}\indicator{(0,\infty)}{(\nu)}$, fixing the value of $h>0$. We have seen in simulations that large values of $h$ tend to make each within-variance, $\sigma_j^2$, to concentrate in the observed sample variance $s_j^2$ at the cost of increasing the uncertainty in their estimates. Meanwhile, smaller values for $h$ have the opposite effect, generating models that are closer to the case where a single common within-variance, $\sigma^2$, is considered for all the groups. However, even while it seems as a reliable approach in the simulations, we do not have any guarantee that the posterior distribution would exist using these improper priors. For example, the limit case $h\to 0$, corresponding with the prior $p(\nu)=\indicator{(0,\infty)}{(\nu)}$, generates an improper monotonically increasing posterior for $\nu$, which makes all the within-variances to concentrate in a common-variance quantity. In this work we present results for $h=3,2,1$.

Note that performing Gibbs sampler for this approach is still possible. To sample from the distribution of $\nu$, we could use a grid of values and sample them with a probability proportional to the (conditional) posterior of those values. However, this requires the extra-effort of finding an appropriate grid.

\subsubsection{Using a regularizing prior for \texorpdfstring{$\nu$}{nu} }
\label{subsubsec:ExponentialNu}

Because setting the value of $\nu$ to a fix value, $\hat{\nu}$, eliminates the uncertainty on $\nu$, and using an improper distribution does not gives guarantee for the existence of the posterior distribution, the third option that we propose is to use a regularizing prior for $\nu$.

A regularizing prior is a prior distribution whose parameters are learn from the data, which might prevent overfitting \cite{mcelreath2018statistical,lemoine2019moving}. With this approach, we maintain the uncertainty on $\nu$ with the guarantee of the existence of the posterior distribution. In this work, we consider an exponential distribution whose rate parameter is set at $1/\hat{\nu}$, but other distributions might be considered as well.

As commented previously, using Gibbs sampler is still feasible, but requires an extra-effort of finding an appropriate grid of values to sample from. However, we can use other Monte Carlo techniques to simulate from the joint posterior distribution. For this purpose, we used No U-Turn Sampler (NUTS) \cite{hoffman2014no} which is a Hamiltonian Monte Carlo technique \cite{DUANE1987216}, implemented in the library \texttt{PyMC} (formerly \texttt{PyMC3}) \cite{salvatier2016probabilistic}.

\subsection{Comparison of models}
\label{sec:CompModel}
We implemented the discussed models for the variables selected in \Cref{subsec:SelectVar} and calculated the WAIC for each one of them. We present in \Cref{tab:WaicAllModels} these values. We show in bold the lowest value of the WAIC for each variable, which corresponds with the preferred model according with this criterion.

\begin{table}[ht]
    \begin{tabular}{lcccc}
     & No & Complete & Hierarchical & Hierarchical \\
     & Pooling & Pooling & common $\sigma^2$ & fixed $\hat\nu$ \\
     \hline
    Monthly Income & 1382.01 & 1410.13 & 1386.44 & \textbf{1378.58} \\
    Percentage with Formal Debt & -216.66 & -183.14 & -217.39 & -221.02 \\
    Formal Debt & 1795.12 & 1816.79 & \textbf{1790.54} & \textbf{1790.54} \\
    Percentage without Savings & -189.07 & -149.70 & -191.04 & -193.89 \\
    Yearly Savings & 1515.78 & 1519.94 & 1518.12 & \textbf{1512.98} \\
    Smoking & -265.82 & -217.48 & -262.48 & -267.08 \\
    Alcohol Consumption & -244.94 & -223.07 & \textbf{-252.58} & -250.27 \\
    \end{tabular}\bigskip
    
    \begin{tabular}{lcccc}
     & Hierarchical & Hierarchical & Hierarchical & Hierarchical \\
     & Exponential($1/\hat\nu$) & $h=3$ & $h=2$ & $h=1$ \\
     \hline
    Monthly Income & 1379.21 & 1380.30 & 1380.15 & 1380.31 \\
    Percentage with Formal Debt & \textbf{-221.04} & -219.30 & -220.08 & -219.97 \\
    Formal Debt & \textbf{1790.54} & 1792.28 & 1791.51 & 1790.73 \\
    Percentage without Savings & \textbf{-193.98} & -191.80 & -191.86 & -192.60 \\
    Yearly Savings & 1513.16 & 1513.94 & 1513.56 & 1514.96 \\
    Smoking & \textbf{-267.70} & -266.74 & -266.99 & -265.20 \\
    Alcohol Consumption & -250.43 & -247.72 & -249.07 & -252.85 \\
    \end{tabular}
    \caption{WAIC for each one of the models previously discussed and all the selected variable. We show in bold the model with the lowest WAIC value for each variable, being the preferred model according with this criterion.}
    \label{tab:WaicAllModels}
\end{table}

More important that its use for model selection, the WAIC can help us for comparison of models. The objective is not to determine which is the correct model, a statement that is probably false for all the models, especially in the field of social science, but to determine which models can be \textit{potentially feasible to explain the data}.

\begin{figure}[ht]
    \centering
    \includegraphics[width=0.65\textwidth]{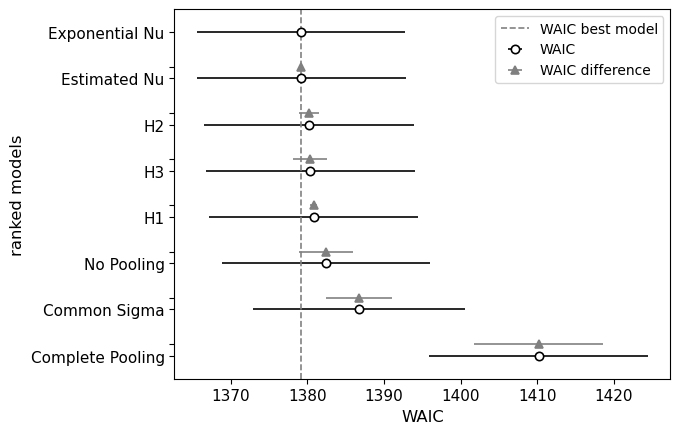}
    \caption{WAIC for the different models implemented for the monthly income variable. The WAIC can hellp us to determine plausible models for the studied phenomenon.}
    \label{fig:WaicCompareModels}
\end{figure}

To answer this problem, a punctual value of the WAIC is not enough. Then, in \Cref{fig:WaicCompareModels}, we present the credible intervals for the WAIC (at a 0.95 posterior probability) of the implemented models for the monthly income variable. The model with the lowest WAIC is when we fix $\nu$ to the estimate value $\hat{\nu}$. However, all the hierarchical models that introduce a multilevel structure in the regional means and the within-region variances are feasible for explaining our data. Meanwhile, the WAIC of the models that assume a common within-variance $\sigma^2$ or complete pooling are far from the preferred one, so we cannot consider them as reliable models for our data.

\begin{figure}[ht]
	\centering
	\begin{minipage}[c]{0.49\textwidth}
		\centering
		\includegraphics[width=\textwidth]{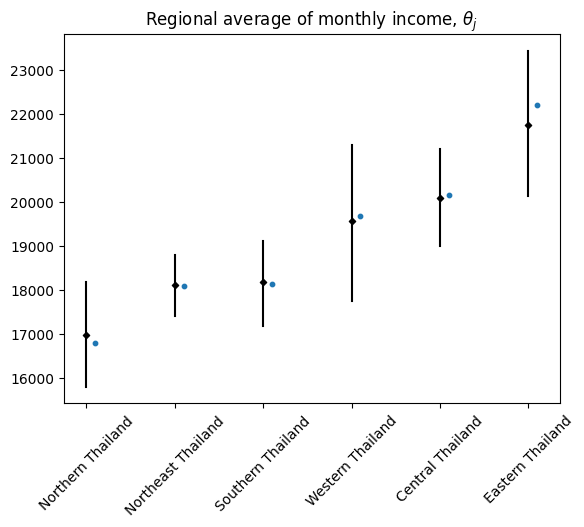}
	\end{minipage}\hfill
	\begin{minipage}[c]{0.49\textwidth}
		\centering
		\includegraphics[width=\textwidth]{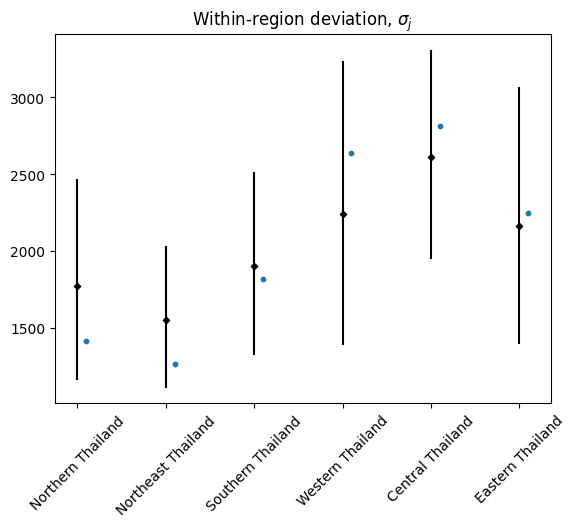}
	\end{minipage}
    \begin{minipage}[c]{0.49\textwidth}
		\centering
		\includegraphics[width=\textwidth]{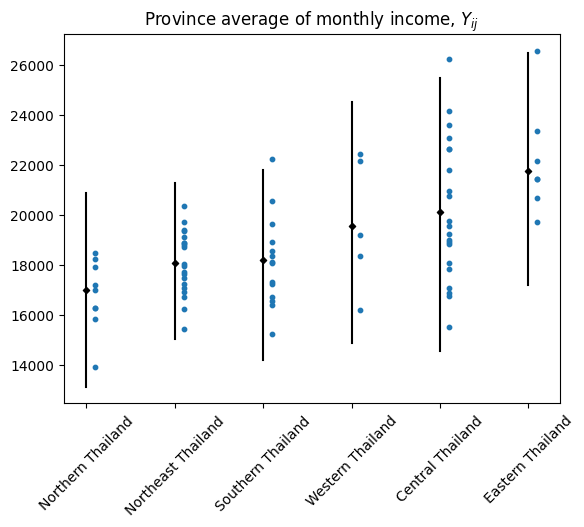}
	\end{minipage}\hfill
	\begin{minipage}[c]{0.49\textwidth}
		\centering
		\includegraphics[width=\textwidth]{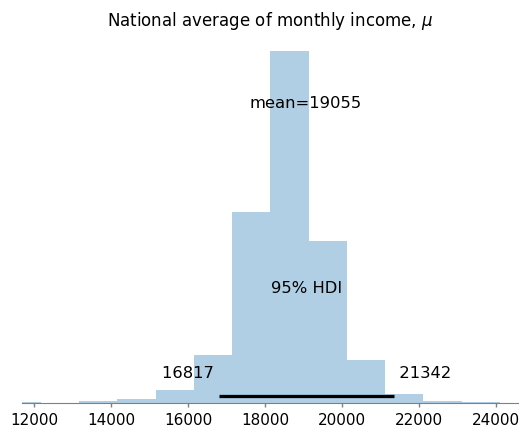}
	\end{minipage}
    \caption{Results considering a hierarchical model. We allow, both the mean of each region and the within-region variance $\sigma_j^2$ to vary. We considered the prior $\sigma_j^2|\nu,\rho^2\sim\textsf{Inverse-}\chi^2(\nu,\rho^2)$. For $\rho^2$ we used the vague prior $p(\rho^2)=\frac{1}{\rho^2}\indicator{(0,\infty)}{(\rho^2)}$, while $\nu$ was fixed to its estimate value $\hat{\nu}$. On the top row we show the regional mean, $\theta_j$ (left) and the regional standard deviation, $\sigma_j$ (right). On the bottom row we show the province mean (left) and the national average monthly income, $\mu$ (right).}
    \label{fig:ResultsHierarchicalFixedNu}
\end{figure}

As it is usual now, \Cref{fig:ResultsHierarchicalFixedNu} shows the results of the model with the lowest WAIC for the monthly income variable. On the left of the top row we present the regional mean of the monthly income, while the deviation within each region is presented on its right. We observe that the model can explain not only the observed average in each region, but also its variability. On the left of the bottom row we show the average monthly income per province in each one of the regions. An important difference from the complete pooling or no pooling models, is that the hierarchical model explicitly add parameters that model the national behavior. For example, the parameter $\mu$ models the national average monthly income, whose posterior distribution is presented on the right of the bottom row in \Cref{fig:ResultsHierarchicalFixedNu}.

\begin{figure}[ht]
	\centering
	\begin{minipage}[c]{0.49\textwidth}
		\centering
		\includegraphics[width=0.85\textwidth]{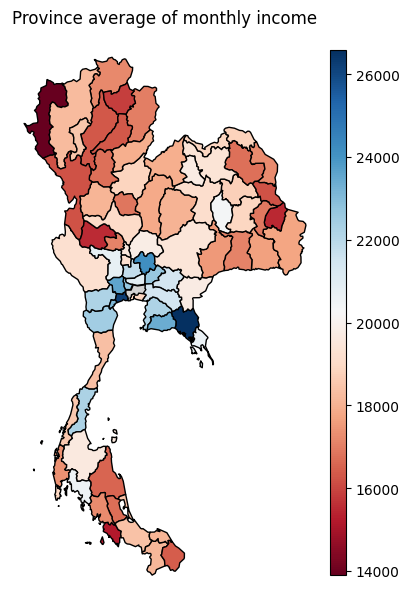}
	\end{minipage}\hfill
	\begin{minipage}[c]{0.45\textwidth}
		\centering
		\includegraphics[width=0.7\textwidth]{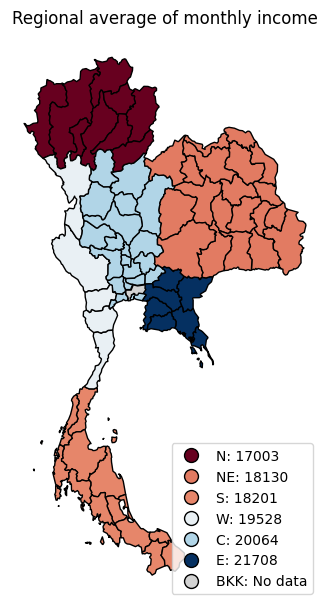}
	\end{minipage}
    \caption{Left: Average monthly income per province. Right: Average regional monthly income.}
    \label{fig:MapPerRegion}
\end{figure}

Finally, \Cref{fig:MapPerRegion} presents two maps of Thailand\footnote{Source for the raw map of Thailand: \href{https://github.com/cvibhagool/thailand-map}{https://github.com/cvibhagool/thailand-map}}, the observed average income per province is presented on the left, while the regional average income is presented on the right.

\section{Hierarchical model with two non-nested clusters: income per region and education level}
\label{sec:HierRegionAndEducation}

\begin{figure}[ht]
\begin{center}
	\begin{tikzpicture}[>=stealth', grow=up,shape=circle,very thick,level distance=16mm,every node/.style={draw=black,circle},edge from parent/.style={<-,draw}]
	\tikzstyle{level 1}=[sibling distance=32mm]
	\tikzstyle{level 2}=[sibling distance=20mm]
	\tikzstyle{level 3}=[sibling distance=12mm]

	\node {$Y_{ijk}$}
		child {node {$\sigma_{jk}^2$}
			child {node {$\rho_k^2$}}
			child {node {$\nu_k$}}
		}
		child {node [rectangle] {$\theta_j+\lambda_k$}
			child {node {$\lambda_k$}
				child {node {$\xi^2$}}
			}
			child {node {$\theta_j$}
				child {node {$\tau^2$}}
				child {node {$\mu$}}			
			}
		};
	\end{tikzpicture}
\caption{Hierarchical model with non-common $\sigma^2$ and two non-nested clusters. Random variables are represented inside circles, while a square represents a deterministic relation.}
\label{fig:HierarchicalMoreClusters}
\end{center}
\end{figure}
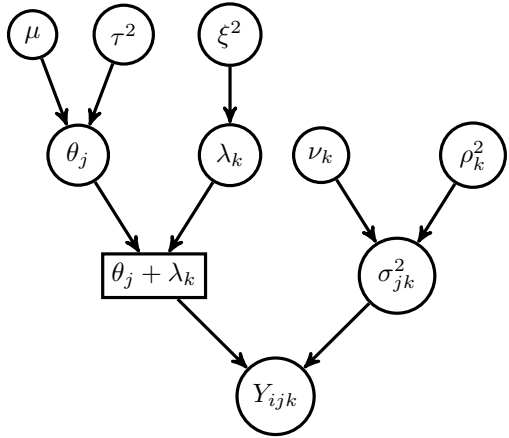

In this section we extend our hierarchical model to estimate the income not only by region, but also by education level. From a statistical perspective adding the education level to our model means that we add another cluster to the model which do not hold a hierarchical structure with the first one. For this purpose, we have assigned the observations to one of three mutually exclusive groups according to the highest education of the people in the house. We have called this variable \textit{education level}, whose possible values are low, mid or high, and whose assignation is done accordingly to the rule presented in \Cref{tab:Education}

In \Cref{tab:ProportionEducation}, we present the percentage of the population belonging to the different levels of education, we can observe that approximately half of the population belongs to the mid education level, i.e. they completed elementary school but do not hold a bachelor or post-graduate degree, with the other two groups representing a significant percentage of the population each one. The percentage of the population with a low education level rounds 27\% while the percentage for those with a high education level is around 22\%. Thus, the amount of observations belonging to each group is large enough to achieve reliable results per education level.

\begin{table}[ht]
\centering
\begin{tabular}{|c|c|c|}
\hline
\textbf{Education} & \textbf{Years of education} & \textbf{Education level} \\
\hline
Uneducated & 0 & \multirow{4}{*}{Low} \\ 
Kindergarten & 0 & \\
Pre-elementary school & 3 & \\
Elementary school & 6 & \\
\hline
Junior high school & 9 & \multirow{3}{*}{Mid} \\
Senior high school & 12 & \\
Vocational degree & 14 & \\
\hline
Bachelor degree & 16 & \multirow{2}{*}{High} \\
Post-graduate & 19 & \\
\hline
\end{tabular}
\caption{Education, education level and years of education.}
\label{tab:Education}
\end{table}

\begin{table}[ht]
\centering
\begin{tabular}{cccc|ccc|cc}
UN-EDU & KDG & P-ELEM & ELEM & JHS & SHS & VD & BD & PG	\\
\hline
0.82\% & 0.03\% & 2.25\% & 23.65\% & 17.35\% & 23.94\% & 9.88\% & 20.97\% & 1.11\% \\
\hline
& & & 26.75\% & & & 51.17\% & & 22.08\%	
\end{tabular}
\caption{Proportion of observations according to the highest education in the house and its corresponding education level.}
\label{tab:ProportionEducation}
\end{table}

Let be $Y_{ijk}$ the average income in province the $i$, belonging to region $j$, when the education level is equal to $k$. We maintain the hierarchical structure for both the regional mean and the within-region variance. But in this case, we allow the variance, $\sigma^2_{jk}$, to vary not only between regions but also between education levels. We now present our hierarchical model, whose Bayesian network is shown in \Cref{fig:HierarchicalMoreClusters}:
\begin{align*}
    Y_{ijk}|\theta_j,\lambda_k,\sigma^2_{jk} &\sim \Normal(\theta_j+\lambda_k,\sigma^2_{jk}) \\
    \theta_j|\mu,\tau^2 &\sim \Normal(\mu,\tau^2) \\
    \mu &\sim \Normal(\hat{\mu}, \hat{\sigma}^2_\mu) \\
    \tau^2 &\sim \textsf{Exponential}(1/\hat{\tau}^2) \\
    \lambda_k|\xi^2 &\sim \Normal(0,\xi^2) \\
    \xi^2 &\sim \textsf{Exponential}(1/\hat{\xi}^2) \\
    \sigma^2_{jk}|\nu_k,\rho^2_k &\sim \textsf{Inverse-}\chi^2(\nu_k,\rho_k^2) \\
    \nu_k^2 &\sim \textsf{Exponential}(1/\hat{\nu}_k^2) \\
    p(\rho_k^2) &\propto \frac{1}{\rho_k^2}\indicator{(0,\infty)}{(\rho_k^2)}
\end{align*}
We proceed to explain the different parts of this model and how the hyperprior distributions where setting.

\paragraph{Likelihood.} The first line of our model corresponds with the likelihood. We model $Y_{ijk}$ as a normal variable with variance $\sigma^2_{jk}$, and mean $\theta_j+\lambda_k$. The average monthly income of region $j$ is still modeled by $\theta_j$, while $\lambda_k$ is interpreted as the additional income due to the education level.

\paragraph{Prior distribution for $\theta_j$.} For the regional average monthly income we use a normal distribution as before, with mean $\mu$ representing the average national monthly income, and variance $\tau^2$ representing the variance of the monthly income between regions.

\paragraph{Prior distribution for $\mu$.} To establish the prior distribution for $\mu$ we use the hierarchical model proposed in \Cref{subsec:HierVaryingSigma}. Then, we use the posterior sample of $\mu$, being $\hat{\mu}$ its average and $\hat{\sigma}^2_\mu$ its variance.

\paragraph{Prior distribution for $\tau^2$.} Similarly to the prior distribution of $\mu$. To establish the prior distribution for $\tau^2$ we use one more time the hierarchical model proposed in \Cref{subsec:HierVaryingSigma}. Then, we use the posterior sample of $\tau$, being $\hat{\tau}$ its average. Note that, in this way, we use previous simpler models as building blocks to construct the priors of more complex models.

\paragraph{Prior distribution for $\lambda_k$.} Consider the mean of $Y_{ijk}$, $\theta_j+\lambda_k$, and note that because $\theta_j|\mu,\tau^2$ and $\lambda_k|\xi^2$ follow normal distributions, it can be written as
\[\theta_j+\lambda_k = \mu+\tau Z_1 + \xi Z_2,\] where $Z_1$ and $Z_2$ are independent standard normal variables. From this expression, it is easy to observe that we have set the mean of $\lambda_k$ to zero to have an identifiable model. If, on the other hand, we introduce a non-zero mean for $\lambda_k$, we would not have any way to distinguish between both $\mu$ and this new hyperparameter.

For a fix $k$, we can estimate $\lambda_k$ as follows. We first implement the hierarchical model proposed in \Cref{subsec:HierVaryingSigma} but only for those observations whose education level equals $k$, let be $\mu_k$ the national monthly income for this model. On the other hand, we implement the same hierarchical model for all the observations (note that this is the model used for the prior specification of both $\mu$ and $\tau^2$). Thus, a punctual estimator for $\lambda_k$, denoted as $\hat{\lambda}_k$, is given by the posterior mean of the variable $\lambda_k=\mu_k-\mu$.

\paragraph{Prior distribution for $\xi^2$.} Because $\xi^2$ represents the variance of $\lambda_1,\ldots,\lambda_K$, we can estimate it with the variance of $\hat{\lambda}_1,\ldots\hat{\lambda}_K$, denoted as $\hat{\xi}^2$. Then, for the prior distribution of $\xi^2$ we use an exponential distribution with rate $1/\hat{\xi}^2$.

\paragraph{Prior distribution for $\sigma_{jk}^2$.} For this model, we assume that the within-region variance can vary not only between regions but also between education levels. For the prior of $\sigma_{jk}^2$ we use the usual inverse-$\chi^2$ distribution presented in \Cref{subsec:HierVaryingSigma}.

\paragraph{Prior distribution for $\nu_k$.} For a fix $k$, we estimate $\nu_k$ through \Cref{eq:HatNu} considering those observations whose education level equals $k$. An exponential distribution with rate parameter equal to $1/\hat{\nu}_k$ is used as proposed in \Cref{subsubsec:ExponentialNu}.

\paragraph{Prior distribution for $\rho_k^2$.} For $\rho_k^2$, we use the vague prior $p(\rho_k^2)\propto\frac{1}{\rho_k^2}\indicator{(0,\infty)}{(\rho_k^2)}$, which yields a proper posterior distribution.

We show in \Cref{tab:WaicEducationLevel} the WAIC with and without considering the education level. We observe that considering the education level leads to a huge reduction of the WAIC, preferring the model that incorporates both clusters. Note that the values of the WAIC are around three times those presented in \Cref{tab:WaicAllModels} for the monthly income variable. However, these quantities are not comparable. The reason is that when the education level is considered we have the union of three datasets, one for each education level for the 76 provinces. Then, the model which incorporates the education level has three times the number of observations of the models that do not incorporate it, making the WAIC incomparable between both set of models.

\begin{table}[ht]
\centering
\begin{tabular}{lcc}
 & Without & With region and \\
 & education level & education level \\
 \hline
Monthly Income & 4662.59 & \textbf{4096.79}
\end{tabular}
\caption{WAIC without considering the education level, and when both region and education level are added to the model.}
\label{tab:WaicEducationLevel}
\end{table}

\begin{figure}[ht]
	\centering
	\begin{minipage}[c]{0.49\textwidth}
		\centering
		\includegraphics[width=\textwidth]{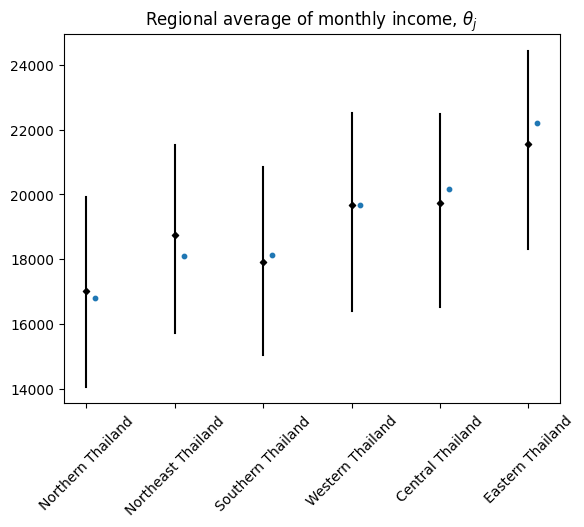}
	\end{minipage}\hfill
	\begin{minipage}[c]{0.49\textwidth}
		\centering
		\includegraphics[width=\textwidth]{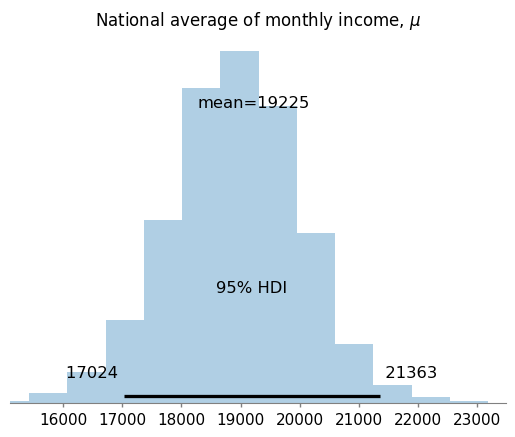}
	\end{minipage}
    \caption{Estimated monthly income, using a hierarchical model that incorporates the region and the education level. Left: Regional average monthly income. Right: National average.}
	\label{fig:MuPerRegionAndEducationLevel}
\end{figure}

\begin{figure}[ht]
	\centering
	\includegraphics[width=\textwidth]{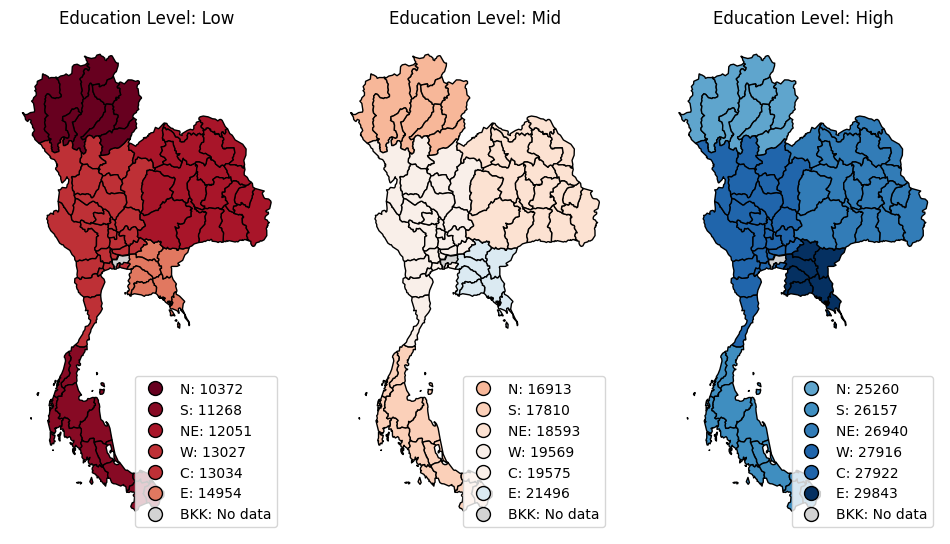}
	\caption{Average monthly income per region and education level.}
	\label{fig:MapMeanIncomePerRegionAndEducationLevel}
\end{figure}

In \Cref{fig:MuPerRegionAndEducationLevel} we present the estimated monthly income at a national (left) and regional level (right). A close inspection reveals a similar distribution for the national average compared with the results presented in \Cref{fig:ResultsHierarchicalFixedNu}. However, the regional averages show more overlap, this indicates that, once we consider the education level, the region has less impact in the income. This might be the result of a mediation relation, in which education level acts as a mediator between the region and the income. This same phenomenon can be observed in \Cref{fig:MapMeanIncomePerRegionAndEducationLevel} with more color-homogeneous maps for each education level, but with large differences between them. For those readers interested in causal inference, we recommend \cite{pearl2000models,pearl2016causal,peters2017elements}.

In \Cref{sec:AdditionalFigures} we present supplementary Figures for this model.

\section{Bayesian hierarchical regression: income considering years of formal education}
\label{sec:HierRegress}

\subsection{National model and separate models}

Instead of considering the education as a categorical variable, we can approximate the years of formal education received, using the rule presented in \Cref{tab:Education}, called this new variable $X$. Then, we can implement a regression model that estimates the income taking as input the years of education.

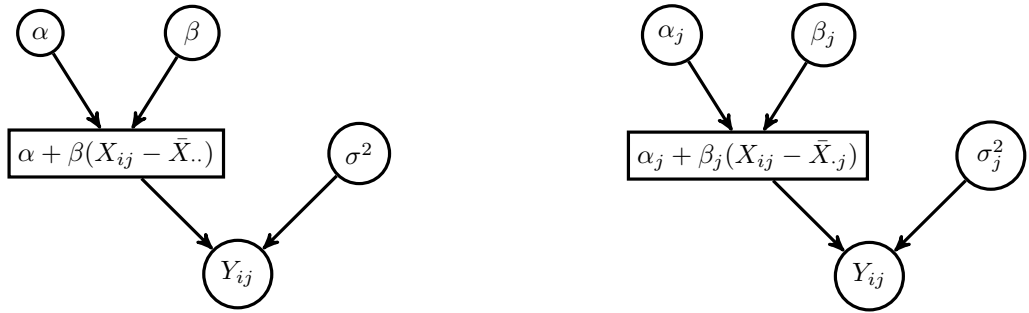
\begin{figure}[ht]
	\centering
	\begin{minipage}[c]{0.45\textwidth}
		\centering
		\begin{tikzpicture}[>=stealth', grow=up,shape=circle,very thick,level distance=16mm,every node/.style={draw=black,circle},edge from parent/.style={<-,draw}]
		\tikzstyle{level 1}=[sibling distance=32mm]
		\tikzstyle{level 2}=[sibling distance=20mm]
		\tikzstyle{level 3}=[sibling distance=12mm]

		\node {$Y_{ij}$}
			child {node {$\sigma^2$}}
			child {node [rectangle] {$\alpha + \beta (X_{ij}-\bar{X}_{\cdot \cdot})$}
				child {node {$\beta$}}
				child {node {$\alpha$}}
			};
		\end{tikzpicture}
	\end{minipage}\hfill
	\begin{minipage}[c]{0.45\textwidth}
		\centering
		\begin{tikzpicture}[>=stealth', grow=up,shape=circle,very thick,level distance=16mm,every node/.style={draw=black,circle},edge from parent/.style={<-,draw}]
		\tikzstyle{level 1}=[sibling distance=32mm]
		\tikzstyle{level 2}=[sibling distance=20mm]
		\tikzstyle{level 3}=[sibling distance=12mm]

		\node {$Y_{ij}$}
			child {node {$\sigma_j^2$}}
			child {node [rectangle] {$\alpha_j + \beta_j (X_{ij}-\bar{X}_{\cdot j})$}
				child {node {$\beta_j$}}
				child {node {$\alpha_j$}}
			};
		\end{tikzpicture}
	\end{minipage}
    \caption{Graphical representation of regression models. Left: National model, where only one regression function is considered for the all the regions. Right: No pooling model, where a regression model is implemented for each region independently of the others.}
	\label{fig:NationalAndIndependentRegressions}
\end{figure}

According to our proposed procedure, before presenting the full hierarchical model, we first consider two simple models. The complete pooling model and separate independent models for each region. These models are represented graphically in \Cref{fig:NationalAndIndependentRegressions}.

\subsubsection{National model}

Let be $X_{ij}$ the average years of formal education in the province $i$ belonging to region $j$. The complete pooling model means the implementation of a single regression function that models the national relation between the income and the years of formal education. With this simple model, we can use vague priors for the parameters without harmful. We introduce now our national model:

\begin{align*}
    Y_{ij}|\alpha,\beta,\sigma^2 &\sim \Normal(\alpha + \beta (X_{ij}-\bar{X}_{\cdot \cdot}),\sigma^2) \\
    p(\alpha,\beta,\sigma^2) &\propto \frac{1}{\sigma^2}\indicator{\R}{(\alpha)}\indicator{\R}{(\beta)}\indicator{(0,\infty)}{(\sigma^2)}
\end{align*}

The expected income $Y_{ij}$ is modeled through the regression function that takes as input the years of education $X_{ij}$, for simplicity we have considered a linear function for the regression, given by $\alpha + \beta (X_{ij}-\bar{X}_{\cdot \cdot})$, where $\bar{X}_{\cdot \cdot}$ is the average years of formal education between all the provinces, that is \[\bar{X}_{\cdot\cdot}=\frac{\sum_{j=1}^J\sum_{i=1}^{n_j}X_{ij}}{n},\] where $n=\sum_{j=1}^J n_j$. Note that $\alpha$ represents the average national income when the years of education equal the national average, while $\beta$ represents the amount of income added per year-of-education.

\begin{figure}[ht]
	\centering
	\includegraphics[width=\textwidth]{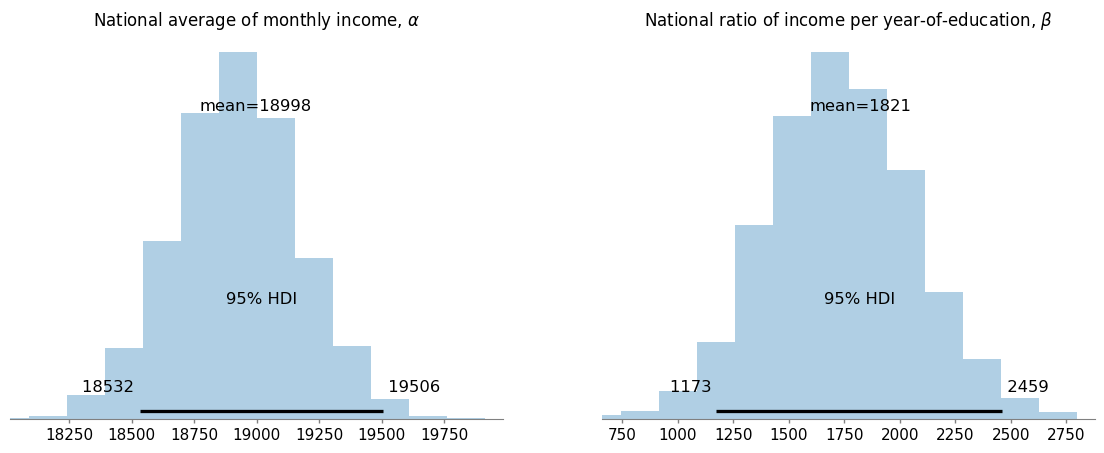}
	\caption{Left: National average monthly income when the years of formal education equal the national average. Right: National ratio income per year-of-education.}
	\label{fig:AlphaBetaCommonAlphaBeta}
\end{figure}

\begin{figure}[ht]
	\centering
	\includegraphics[width=0.65\textwidth]{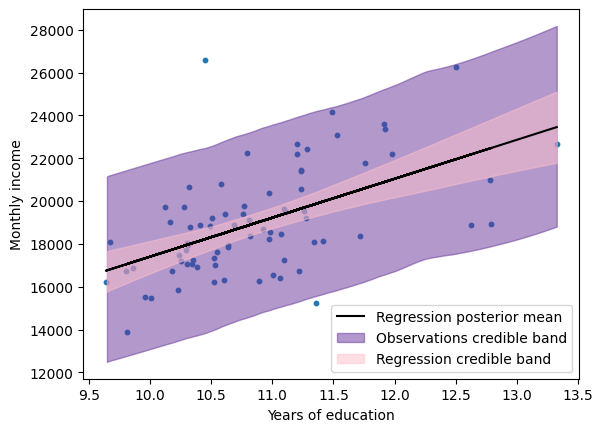}
	\caption{National regression model for the monthly income given the years of formal education.}
	\label{fig:RegressionCommonAlphaBeta}
\end{figure}

We present in \Cref{fig:AlphaBetaCommonAlphaBeta} the posterior distributions for $\alpha$ and $\beta$, \Cref{fig:RegressionCommonAlphaBeta} presents the estimated regression function, we also present credible bands for the regression function and the income, both bands are calculated at a 0.95 posterior probability.

\subsubsection{Separate models}
Instead of considering just one regression function for all the regions, we can estimate a regression function for each one of the regions, independent from the others. Similarly to the national regression model, for each one of these models we can use vague priors for the parameters without harmful:

\begin{align*}
    Y_{ij}|\alpha_j,\beta_j,\sigma_j^2 &\sim \Normal(\alpha_j + \beta_j (X_{ij}-\bar{X}_{\cdot j}),\sigma_j^2) \\
    p(\alpha_j,\beta_j,\sigma_j^2) &\propto \frac{1}{\sigma_j^2}\indicator{\R}{(\alpha_j)}\indicator{\R}{(\beta_j)}\indicator{(0,\infty)}{(\sigma_j^2)}
\end{align*}

We consider a linear function for the regression in each region $j$, given by $\alpha_j + \beta_j (X_{ij}-\bar{X}_{\cdot j})$, where \[\bar{X}_{\cdot j}=\frac{\sum_{i=1}^{n_j}X_{ij}}{n_j}\] is the average years of formal education in the region $j$. Note that $\alpha_j$ represents the average income in the region when the years of education equal the regional average, while $\beta_j$ represents the amount of income added per year-of-education in the region.

\begin{figure}[ht]
	\centering
	\includegraphics[width=\textwidth]{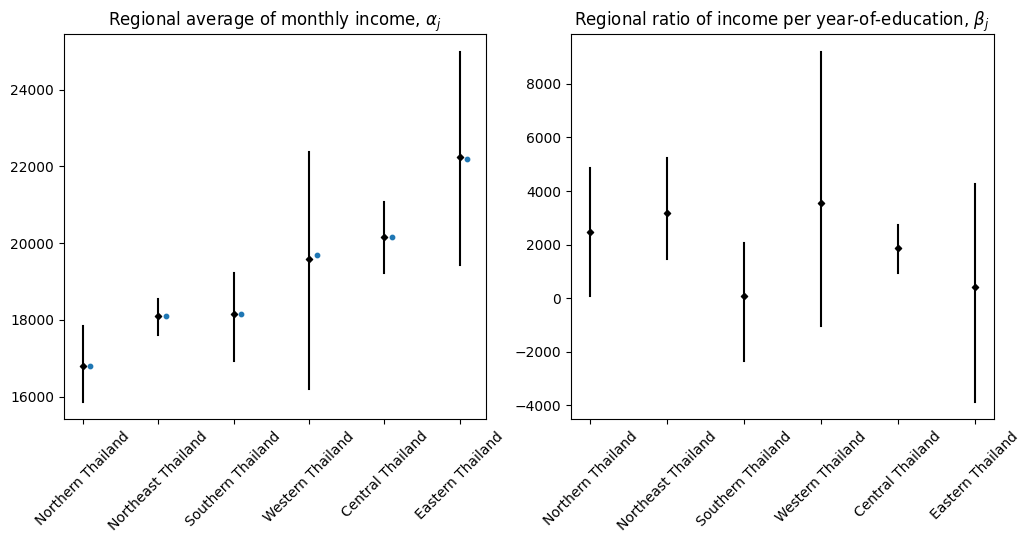}
	\caption{Left: Regional average monthly income when the years of formal study equal the regional average. Right: Regional ratio income per year-of-education.}
	\label{fig:AlphaBetaSeparateAlphaSeparateBeta}
\end{figure}

On the left of \Cref{fig:AlphaBetaSeparateAlphaSeparateBeta}, we present the average monthly income for each region when the years of education are equal to the regional mean, $\alpha_j$, with their respective credible intervals of 0.95 posterior probability and the observed average income in each region. Analogously, on the right we present the ratio of income per year-of-education, $\beta_j$. Because this model assumes that the regions do not share any information, we observe large credible intervals, for some regions like Southern Thailand or Eastern Thailand these intervals even include negative values, which seems implausible. Moreover, as we pointed before, pretending that each region is independent for the others seems unrealistic. Therefore, we introduce the hierarchical regression model as a compromise between a single regression model and an independent regression model for each region.

\subsection{Bayesian hierarchical regression varying intercepts}

\begin{figure}[ht]
\begin{center}
	\begin{tikzpicture}[>=stealth', grow=up,shape=circle,very thick,level distance=16mm,every node/.style={draw=black,circle},edge from parent/.style={<-,draw}]
	\tikzstyle{level 1}=[sibling distance=32mm]
	\tikzstyle{level 2}=[sibling distance=20mm]
	\tikzstyle{level 3}=[sibling distance=12mm]

	\node {$Y_{ij}$}
		child {node {$\sigma_j^2$}
			child {node {$\rho^2$}}
			child {node {$\nu$}}
		}
		child {node [rectangle] {$\alpha_j + \beta (X_{ij}-\bar{X}_{\cdot j})$}
			child {node {$\beta$}}
			child {node {$\alpha_j$}
				child {node {$\tau^2$}}
				child {node {$\mu$}}
			}
		};
	\end{tikzpicture}
\caption{Graphical representation of the Bayesian hierarchical regression model, varying intercepts but considering a common slope for all the regions.}
\label{fig:RegressionCommonAlphaSeparateBetaDag}
\end{center}
\end{figure}
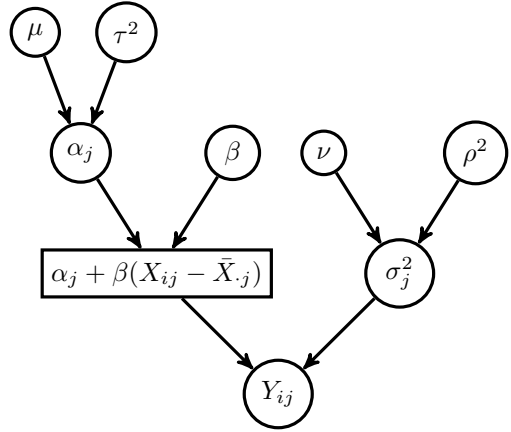

We already observed in \Cref{subsec:HierVaryingSigma} that adding a common structure to the region average income and to the within-regions variance results in a better model, so we can implement a model with those characteristics for the regression task. This model is represented graphically in \Cref{fig:RegressionCommonAlphaSeparateBetaDag}.

We now present our hierarchical model for the regression task:
\begin{align*}
    Y_{ij}|\alpha_j,\beta_j,\sigma_j^2 &\sim \Normal(\alpha_j + \beta_j (X_{ij}-\bar{X}_{\cdot j}),\sigma_j^2) \\
    \alpha_j|\mu,\tau^2 &\sim \Normal(\mu,\tau^2) \\
    \mu &\sim \Normal(\hat{\mu}, \hat{\sigma}^2_\mu) \\
    \tau^2 &\sim \textsf{Exponential}(1/\hat{\tau}^2) \\
    p(\beta) &\propto \indicator{\R}{(\beta)} \\
    \sigma^2_j|\nu,\rho^2 &\sim \textsf{Inverse-}\chi^2(\nu,\rho^2) \\
    \nu^2 &\sim \textsf{Exponential}(1/\hat{\nu}^2) \\
    p(\rho^2) &\propto \frac{1}{\rho^2}\indicator{(0,\infty)}{(\rho^2)}
\end{align*}

Many parts of this model have been inherited from our previous hierarchical models. Then, we explain only the prior distributions of the hyperparameters that changed from the previous models.

\paragraph{Prior distribution for $\mu$.} Note that $\mu$ has a similar interpretation than the intercept parameter in the national regression model. Thus, we set $\hat{\mu}$ and $\hat{\sigma}_\mu$ to the mean and deviation (respectively) of the intercept parameter in the national regression model.

\paragraph{Prior distribution for $\tau^2$.} Note that $\tau^2$ models the variance between $\alpha_1,\ldots,\alpha_J$. Then, we can estimate this quantity from the no pooling regression model. To do so, we first compute the average for the intercept parameter for each region, and then take the variance between these values.

\begin{figure}[ht]
	\centering
	\begin{minipage}[c]{0.45\textwidth}
		\centering
		\includegraphics[width=0.9\textwidth]{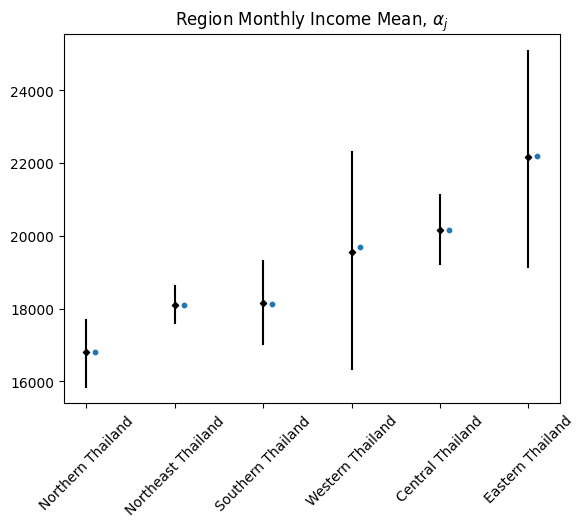}
	\end{minipage}\hfill
	\begin{minipage}[c]{0.45\textwidth}
		\centering
		\includegraphics[width=0.9\textwidth]{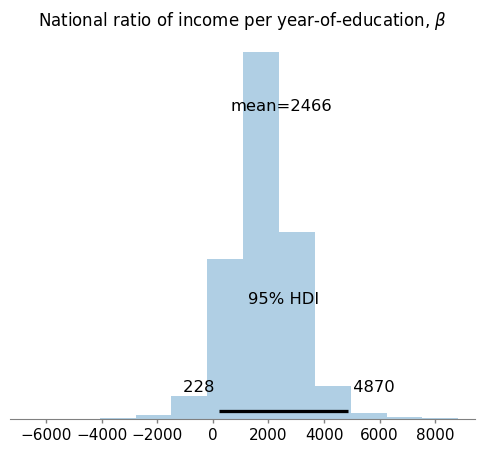}
	\end{minipage}
    \caption{Left: Regional average monthly income when the years of formal study equal the regional average. Right: National ratio income per year-of-education.}
	\label{fig:AlphaBetaVaryingAlphaCommonBeta}
\end{figure}

In \Cref{fig:AlphaBetaVaryingAlphaCommonBeta} we present analogous graphs of those presented in \Cref{fig:AlphaBetaCommonAlphaBeta,fig:AlphaBetaSeparateAlphaSeparateBeta}. Note that the distribution of $\beta$ includes negative values, which could indicate that different slopes should be preferred instead of a common national parameter.

\subsection{Bayesian hierarchical regression varying intercepts and slopes}
\label{subsec:HierRegressVaryAll}

\begin{figure}[ht]
\begin{center}
	\begin{tikzpicture}[>=stealth', grow=up,shape=circle,very thick,level distance=16mm,every node/.style={draw=black,circle},edge from parent/.style={<-,draw}]
	\tikzstyle{level 1}=[sibling distance=32mm]
	\tikzstyle{level 2}=[sibling distance=20mm]
	\tikzstyle{level 3}=[sibling distance=12mm]

	\node {$Y_{ij}$}
		child {node {$\sigma_j^2$}
			child {node {$\rho^2$}}
			child {node {$\nu$}}
		}
		child {node [rectangle] {$\alpha_j + \beta_j (X_{ij}-\bar{X}_{\cdot j})$}
			child {node {$\alpha_j, \beta_j$}
				child {node {$\rho_{\alpha,\beta}$}}
				child {node {$\zeta^2$}}
				child {node {$\gamma$}}			
				child {node {$\tau^2$}}
				child {node {$\mu$}}
			}
		};
	\end{tikzpicture}
\caption{Graphical representation of the hierarchical Bayesian regression model, varying intercepts and slopes.}
\label{fig:RegressionAlphaBetaCorrDag}
\end{center}
\end{figure}
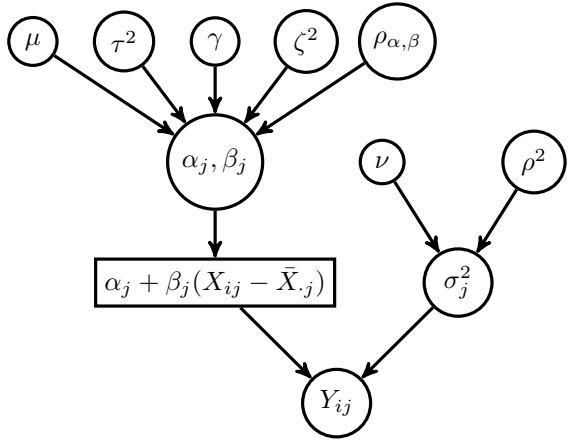

Because a common slope for all the regions seems inappropriate for this case, we now present a model that implements a hierarchical model on both parameters, the intercept and the slope. This model is represented graphically in \Cref{fig:RegressionAlphaBetaCorrDag}.

Instead of just incorporating a normal distribution for the slopes into the previous model, we model the intercepts and slopes through a multivariate normal distribution, allowing them to covary. Then, $\alpha_j$ and $\beta_j$ will follow a multivariate normal distribution with mean $(\mu,\gamma)$ and a matrix of variances and covariances
\[S =
\begin{pmatrix}
    \tau^2 & \tau\zeta\rho_{\alpha,\beta} \\
    \tau\zeta\rho_{\alpha,\beta} & \zeta^2
\end{pmatrix},\]
which can be written as
\[S=\begin{pmatrix}
    \tau & 0 \\
    0 & \zeta
\end{pmatrix} R 
\begin{pmatrix}
    \tau & 0 \\
    0 & \zeta
\end{pmatrix},\]
where
\[R=\begin{pmatrix}
    1 & \rho_{\alpha,\beta} \\
    \rho_{\alpha,\beta} & 1
\end{pmatrix}\]
is the correlation matrix.

We are now ready to present our hierarchical model:
\begin{align*}
    Y_{ij}|\alpha_j,\beta_j,\sigma_j &\sim \textsf{Laplace}(\alpha_j + \beta_j (X_{ij}-\bar{X}_{\cdot j}),\sigma_j) \\
    \alpha_j,\beta_j|\mu,\tau^2,\gamma,\zeta^2,\rho_{\alpha,\beta} &\sim MVN\parent{\begin{bmatrix} \mu \\ \gamma\end{bmatrix}, S} \\
    \mu &\sim \Normal(\hat{\mu}, \hat{\sigma}^2_\mu) \\
    \tau^2 &\sim \textsf{Exponential}(1/\hat{\tau}^2) \\
    \gamma &\sim \Normal(\hat{\gamma}, \hat{\sigma}^2_\gamma) \\
    \zeta^2 &\sim \textsf{Exponential}(1/\hat{\zeta}^2) \\
    R &\sim \textsf{LKJ}(2) \\
    \sigma^2_j|\nu,\rho^2 &\sim \textsf{Inverse-}\chi^2(\nu,\rho^2) \\
    \nu^2 &\sim \textsf{Exponential}(1/\hat{\nu}^2) \\
    p(\rho^2) &\propto \frac{1}{\rho^2}\indicator{(0,\infty)}{(\rho^2)}
\end{align*}

\paragraph{Likelihood.} In this model we do not consider anymore a normal likelihood for our data. Instead, we consider a Laplace distribution. The density of a random variable $Y$ that follows a Laplace distribution with parameters $\mu$ and $\sigma$ is given by \[p(Y|\mu,\sigma)=\frac{1}{2\sigma}\exp\parent{-\frac{|Y-\mu|}{\sigma}}.\] This change in the likelihood is analogous to median regression in which the absolute errors are minimized, and thus corresponding to a robust regression model. As mentioned in \cite{jeffreynotes}, this can be generalized to other quantiles using the asymmetric Laplace distribution \cite{benoit2017bayesqr,yu2005three}.

\begin{figure}[ht]
	\centering
	\includegraphics[width=\textwidth]{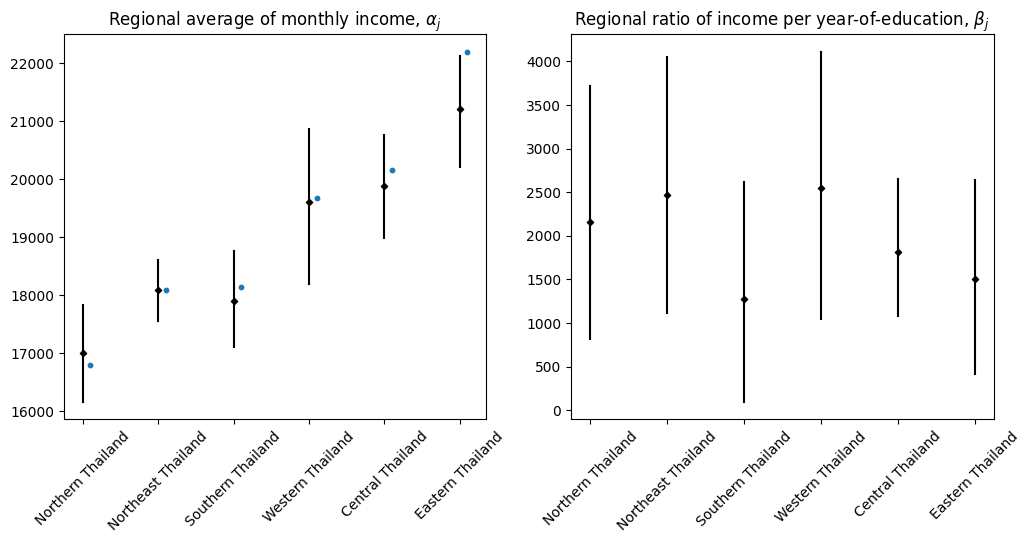}
	\caption{Left: Regional average monthly income when the years of formal study equal the regional average. Right: Regional ratio income per year-of-education.}
	\label{fig:AlphaBetaVaryingAlphaVaryingBeta}
\end{figure}

We can observe in \Cref{fig:AlphaBetaVaryingAlphaVaryingBeta} and \Cref{tab:IncomePerEducation} that with this change none of the credible intervals (calculated at a 0.95 posterior probability) for the ratio of income per year-of-education includes negative values, even while a constraint of positive values for these parameters was not incorporated in the model. If instead of a Laplace distribution, we would have considered a normal distribution for the data, some of these intervals would include negative values, which seems unfeasible to explain.

\paragraph{Prior distribution for $\gamma$.} Analogously to the prior distribution for $\mu$, we set $\hat{\gamma}$ and $\hat{\sigma}^2_\gamma$ to the mean and variance of the posterior distribution for the slope parameter in the national regression model.

\paragraph{Prior distribution for $\zeta^2$.} To set the prior of $\zeta^2$ we follow the same strategy used for the prior of $\tau^2$. That is, we calculate the posterior mean of the slopes for the separate independent regression models, and then we set $\hat{\zeta}^2$ to the variance of these posterior means.

\paragraph{Prior distribution for $R$.} For the prior of $R$ we consider the LKJ distribution \cite{lewandowski2009generating}, which is a distribution over all positive definite correlation matrices where the shape is determined by a single parameter, $\eta>0$. Setting $\eta=1$ results in a uniform distribution over all the correlations $\rho_{\alpha,\beta}$, however setting $\eta=2$ is an alternative that has been considered in the literature \cite{mcelreath2018statistical,wang2018equivalence,follett2020explaining,sorensen2015bayesian} to define a weakly informative prior over $\rho_{\alpha,\beta}$. It implies that $\rho_{\alpha,\beta}$ is near zero, reflecting the prior belief that there is no correlation between intercepts and slopes.

In \Cref{fig:AlphaBetaVaryingAlphaVaryingBeta} we present analogous graphs of those presented in \Cref{fig:AlphaBetaCommonAlphaBeta}. Comparing the credible interval of the slopes, we can observe not only narrower intervals, but also that they contain only positive values, which means that every extra year of education generates a higher income. The credible intervals for the slopes and their posterior mean are also shown in \Cref{tab:IncomePerEducation}. The fact that these intervals overlap suggests that this increment in the income is similar for all the regions.

\begin{figure}[ht]
	\centering
	\includegraphics[width=\textwidth]{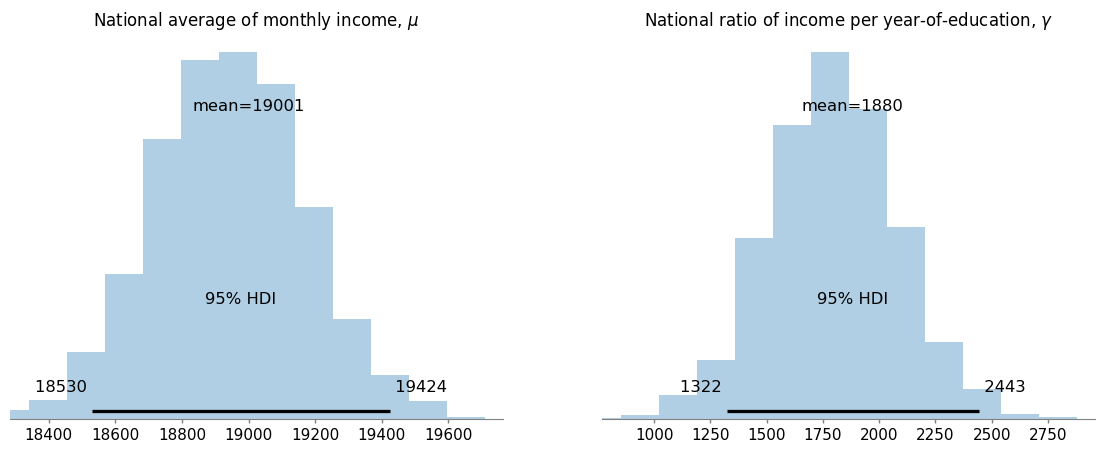}
	\caption{Left: National average monthly income. Right: National ratio of income per-year-of-education}
	\label{fig:MuGammaVaryingAlphaVaryingBeta}
\end{figure}

With our hierarchical model, we can estimate the national average income when the years of education equal the national average years, which is modeled by $\mu$, and whose posterior distribution is shown on the left of \Cref{fig:MuGammaVaryingAlphaVaryingBeta}. On the right side we present the posterior distribution of $\gamma$, which models the national ratio of income-per-year of education.

\begin{table}[ht]
    \centering
    \begin{tabular}{l|c}
         & Amount of monthly income \\
         & added per year-of-education (THB) \\
        \hline
        National level & 1880; (1322, 2443) \\
        \hline
        Northern Thailand & 2159; (814, 3719) \\
        Southern Thailand & 2466; (1111, 4056) \\
        Western Thailand & 1278; (94, 2613) \\
        Eastern Thailand & 2545; (1051, 4109) \\
        Northeast Thailand & 1818; (1076, 2648) \\
        Central Thailand & 1509; (413, 2642)
    \end{tabular}
    \caption{Amount of monthly income added per each year-of-education. We present the posterior mean for the national level and for each one of the regions, as well as a credible interval of 0.95 posterior probability.}
    \label{tab:IncomePerEducation}
\end{table}

\begin{figure}[H]
	\centering
	\includegraphics[width=\textwidth]{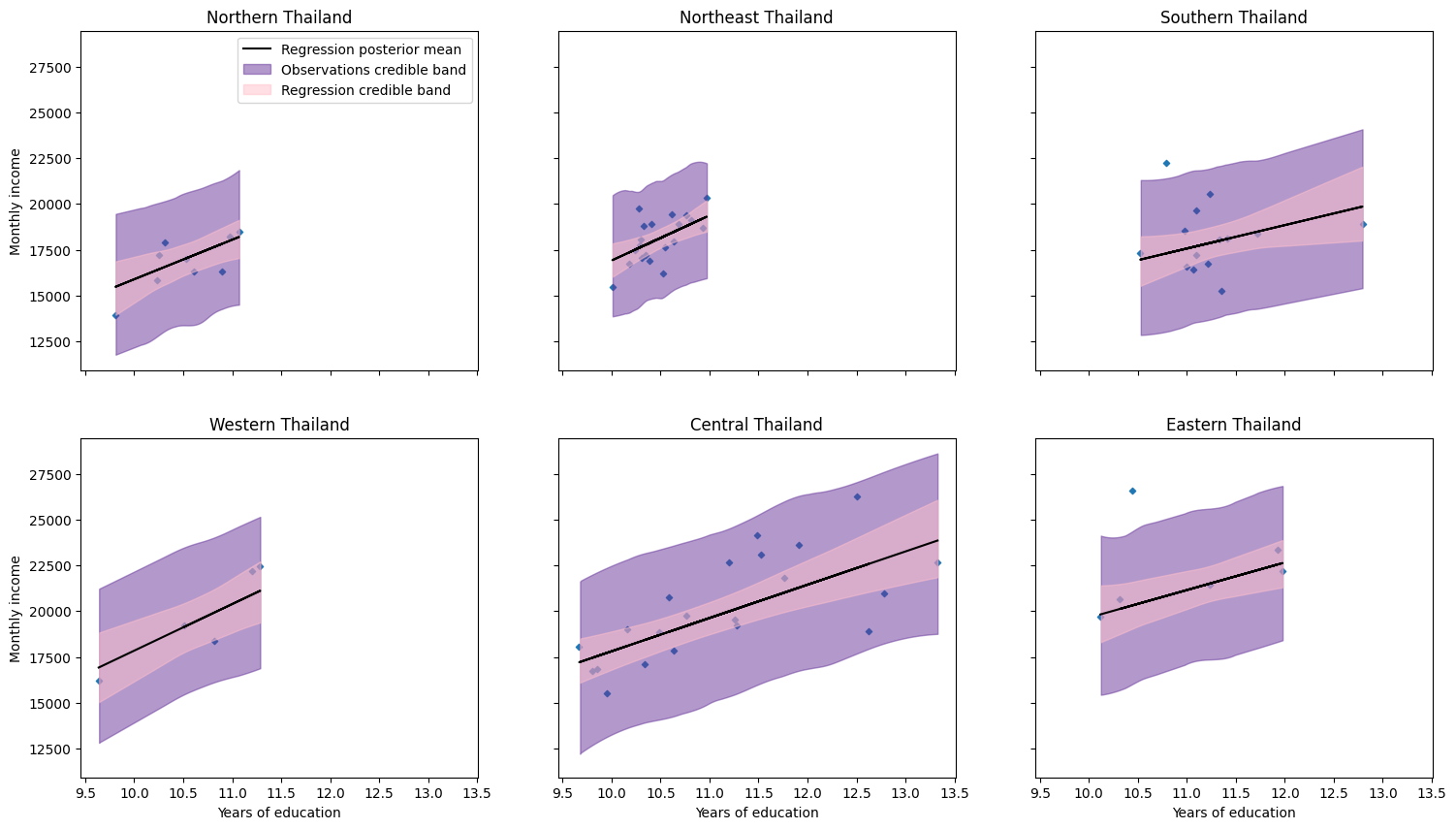}
	\caption{Regional regression models for the monthly income given the years of education, we impose a hierarchical model for the intercepts and the slopes, allowing them to covary.}
	\label{fig:RegressionVaryingAlphaVaryingBeta}
\end{figure}

\begin{figure}[ht]
	\centering
	\begin{minipage}[c]{0.45\textwidth}
		\centering
		\includegraphics[width=0.9\textwidth]{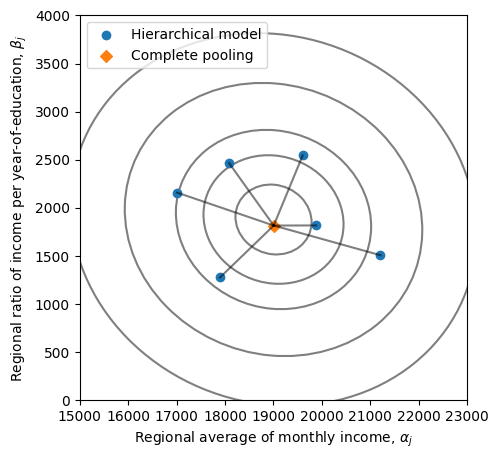}
	\end{minipage}\hfill
	\begin{minipage}[c]{0.45\textwidth}
		\centering
		\includegraphics[width=0.9\textwidth]{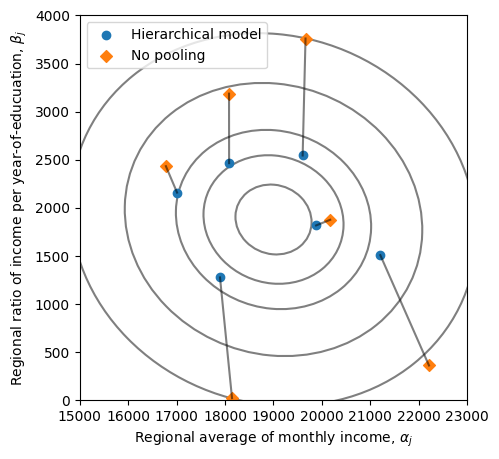}
	\end{minipage}
 	\caption{Joint posterior of the slopes and intercepts. Left: We show the posterior mean of each pair $(\alpha_j,\beta_j)$ and compare them to the posterior mean of the pair $(\alpha,\beta)$ for the national regression model. Right: We present our estimates for the intercepts and slopes and compare them with the estimators for separate independent models for each region.}
	\label{fig:AlphaBetaVsCommonVsSeparate}
\end{figure}

We show in \Cref{fig:RegressionVaryingAlphaVaryingBeta} the estimated regression models with credible bands for the regression functions and the province monthly income, both bands are calculated at a 0.95 posterior probability. In \Cref{fig:AlphaBetaVsCommonVsSeparate} we show the joint posterior of the intercepts and slopes. On the left side we show the posterior mean of each pair $(\alpha_j,\beta_j)$ and compare them to the posterior mean of the pair $(\alpha,\beta)$ for the national regression model. Similarly, on the right side we present our estimates for the slopes and intercepts, and compare them with the extreme case of considering separate independent models for each region. We can observe how the estimators are closer to more probable regions when we impose a hierarchical structure.

\begin{figure}[ht]
    \centering
    \includegraphics[width=0.65\textwidth]{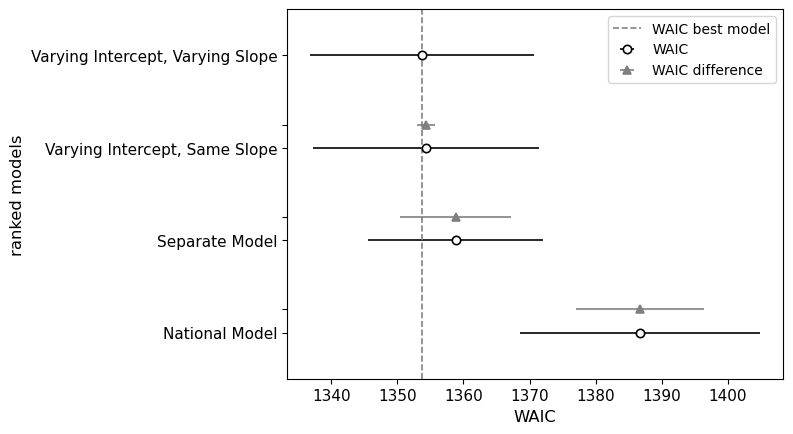}
    \caption{WAIC for the Bayesian hierarchical regression models.}
    \label{fig:WaicRegress}
\end{figure}

Finally, we show in \Cref{tab:WaicRegression} and \Cref{fig:WaicRegress} the WAIC for all the regression models. We also present in \Cref{tab:WaicRegression} the WAIC for the preferred model without a covariate variable (see \Cref{tab:WaicAllModels}). Note that the model that allows variation between intercepts with a common slope is as feasible as the model that allows different intercepts and slopes. We observe that, except for the national regression model, all the regression models present a lower WAIC value, making them more reliable models accordingly to this criterion.

\begin{table}[ht]
\centering
\begin{tabular}{lccccc}
& Without & National & Separate & Varying $\alpha_j$, & Varying $\alpha_j$, \\
& Covariable & Model & Models & Common $\beta$ & Varying $\beta_j$ \\
 \hline
Monthly Income & 1378.58 & 1387.08 & 1358.77 & 1354.37 & \textbf{1353.82}
\end{tabular}
\caption{WAIC for the models of regression. In the first cell we present the lowest value in \Cref{tab:WaicAllModels} for the models without a covariate}
\label{tab:WaicRegression}
\end{table}

\section{Discussion of results and conclusions}
\label{sec:Conclusions}

Throughout this work we presented and discussed several Bayesian hierarchical models. These models were implemented for the variables with the largest percentage of Thai households' affected. Instead of starting with complex hierarchical models, we first introduced the extreme cases of no pooling and complete pooling models, corresponding to one-shirt-size and custom-made policies, respectively. After analyzing the results of these models and their lack of explanation for certain aspects of the data, the hierarchical models were presented as a better approach, being a trade-off between these two extremes.

In \Cref{subsec:HierVaryingSigma}, we extended our previous hierarchical model, incorporating a multilevel structure for the within-cluster variance through an inverse $\chi^2$ distribution, and proposed three different approaches for the prior specification of the degrees of freedom for this distribution. For the comparison of the models, alongside the analysis and discussion of the results, we used the Widely Applicable Information Criterion (WAIC), which always preferred the hierarchical models for the variables considered. Later, in \Cref{sec:HierRegionAndEducation}, we introduced the education level into our hierarchical model, creating a model with two non-nested clusters. \Cref{sec:HierRegress} was devoted to the discussion of Bayesian hierarchical regression, explaining the households' monthly income as a function of the years of education. We achieved the best performance when a robust hierarchical regression was implemented.

From these analyses, we could observe the important impact of the education in the income, always showing a positive relation, and which almost vanished the effect of the region in the income. Furthermore, we were able to estimate the average income for each education level and the ratio of income per year-of-education, at a regional and national levels.

Throughout this work we explained how simple models can be used as building blocks for the prior specification of the hyperparameters in more complex hierarchical models.
 
\section*{Note}
Codes to reproduce our results are available in \href{https://github.com/IrvingGomez/BayesianHierarchicalIncome}{https://github.com/IrvingGomez/BayesianHierarchicalIncome}

\section*{Acknowledgement}
The authors would like to give a special thanks to Dr. Anon Plangprasopchok and Ms. Kittiya Ku-kiattikun who managed the data, cleaned it, and provided insight about the data for us. 

\bibliographystyle{unsrt}
\bibliography{mainBIB}

\begin{appendix}
\section{Technical Appendix}
\label{sec:TechnicalAppendix}

\subsection{WAIC}
\label{subsec:WAIC}

We can compare the average log-probability for each model to get an estimate of the relative distance of each model from the real distribution of the data. The Bayesian version of the log-probability score is called the log-pointwise-predictive-density (lppd) defined as \[\text{lppd}=\sum_{i=1}^n\log\corchet{\frac{1}{S}\sum_{s=1}^S p\parent{Y_i|\Theta^{(s)}}},\] where $S$ is the number of simulated samples from the posterior distributions, $\Theta^{(s)}$ is the $s$-th set of sampled parameter values and $p\parent{Y_i|\Theta^{(s)}}$ is the density of $Y_i$ given the parameters $\Theta^{(s)}$.

However, we should also consider the `complexity' of the model, this complexity is given by the effective number of parameters of the model, labeled $p_{\text{WAIC}}$ and defined as \[p_{\text{WAIC}}=\sum_{i=1}^n\V_{\Theta\sim p(\Theta|\Y)}\log p(Y_i|\Theta).\] Because we count with a sample of the posterior distribution of the parameters, this quantity can be well-approximated taking the sample variance of the log-density for each observation $i$, and then summing up these variances.

Therefore, to get a quantitative way to compare the models, we consider the Watanabe-Akaike information criterion (WAIC) defined as \[\text{WAIC}=-2\text{lppd}+2p_{\text{WAIC}}.\]

In Watanabe’s original definition, WAIC is the negative of the average $\text{lppd}$, thus is divided by $n$ and does not have the factor 2; it is common to scale it to be comparable with AIC and other measures of deviance.

\subsection{Qualitative characteristics of the parameters and models}
\label{Subsec:AppQuali}

Consider the model presented in \Cref{subsec:HierCommonSigma}, and the prior $p(\mu)\propto\indicator{\R}{(\mu)}$, it is not difficult to prove \cite{gelman2013bayesian} that \[\mu|\tau,\sigma^2,\Y\sim\Normal(\hat{\mu}, V_\mu),\]
where
\[\hat{\mu} = \frac{\sum_{j=1}^J \frac{\bar{Y}_{\cdot j}}{\bar\sigma_j^2+\tau^2}}{\sum_{j=1}^J \frac{1}{\bar\sigma_j^2+\tau^2}},\quad V_\mu = \frac{1}{\sum_{j=1}^J \frac{1}{\bar\sigma_j^2+\tau^2}}\text{, and}\quad \bar\sigma_j^2=\frac{\sigma^2}{n_j}.\]
Note that
\[\hat{\mu}\xrightarrow[\tau\to 0]{}\frac{\sum_{j=1}^J \frac{\bar{Y}_{\cdot j}}{\bar\sigma_j^2}}{\sum_{j=1}^J \frac{1}{\bar\sigma_j^2}}\equiv \bar{Y}_{\cdot\cdot}\text{ and}\quad V_\mu\xrightarrow[\tau\to 0]{}\frac{1}{\sum_{j=1}^J \frac{1}{\bar\sigma_j^2}}\equiv\varphi^2.\]

Using these expressions and \Cref{eq:HatThetaAndVTheta}, we can calculate
\begin{align}
\E(\theta_j|\tau,\sigma^2,\Y) & = \E_\mu[\E(\theta_j|\mu,\tau^2,\sigma^2,\Y)|\tau,\sigma^2,\Y] \nonumber\\
& = \E_\mu[\hat{\theta}_j|\tau,\sigma^2,\Y] \nonumber\\
& = \E_\mu\corchet{\frac{\frac{1}{\bar\sigma_j^2}\bar{Y}_{\cdot j}+\frac{1}{\tau^2}\mu}{\frac{1}{\bar\sigma_j^2}+\frac{1}{\tau^2}}\Big|\tau,\sigma^2,\Y} \nonumber\\
& = \frac{\frac{1}{\bar\sigma_j^2}\bar{Y}_{\cdot j}+\frac{1}{\tau^2}\hat{\mu}}{\frac{1}{\bar\sigma_j^2}+\frac{1}{\tau^2}},
\label{eq:ETheta}
\end{align}
and
\begin{align*}
\V(\theta_j|\tau,\sigma^2,\Y) & = \E_\mu[\V(\theta_j|\mu,\tau^2,\sigma^2,\Y)|\tau,\sigma^2,\Y] + \V_\mu[\E(\theta_j|\mu,\tau^2,\sigma^2,\Y)|\tau,\sigma^2,\Y] \\
& = \E_\mu[V_{\theta_j}|\tau,\sigma^2,\Y] + \V_\mu[\hat{\theta}_j|\tau,\sigma^2,\Y] \\
& = \E_\mu\corchet{\frac{1}{\frac{1}{\bar\sigma_j^2}+\frac{1}{\tau^2}}\Big|\tau,\sigma^2,\Y} + \V_\mu\corchet{\frac{\frac{1}{\bar\sigma_j^2}\bar{Y}_{\cdot j}+\frac{1}{\tau^2}\mu}{\frac{1}{\bar\sigma_j^2}+\frac{1}{\tau^2}}\Big|\tau,\sigma^2,\Y} \\
& = \frac{1}{\frac{1}{\bar\sigma_j^2}+\frac{1}{\tau^2}} + \frac{\parent{\frac{1}{\tau^2}}^2 V_\mu}{\parent{\frac{1}{\bar\sigma_j^2}+\frac{1}{\tau^2}}^2}
\end{align*}

Note that $\E(\theta_j|\tau,\sigma^2,\Y)\xrightarrow[\tau\to 0]{}\bar{Y}_{\cdot\cdot}$, and $\V(\theta_j|\tau,\sigma^2,\Y)\xrightarrow[\tau\to 0]{}\varphi^2$. Meanwhile, $\E(\theta_j|\tau,\sigma^2,\Y)\xrightarrow[\tau\to \infty]{}\bar{Y}_{\cdot j}$, and $\V(\theta_j|\tau,\sigma^2,\Y)\xrightarrow[\tau\to \infty]{}\bar\sigma_j^2$. This implies that both the complete pooling and no pooling models can be seen as extreme cases of the hierarchical model, achieved when $\tau\to 0$ and $\tau\to\infty$, respectively. Then, the hierarchical model provides a compromise between these two extreme models.

On the other hand, in \cite{gelman2013bayesian} it is discussed an empirical approach, based on an analysis of variance (ANOVA), to estimate the parameters $\sigma^2$ and $\tau^2$, which we now present and analyze here.

The mean square within groups $MS_W$ is given by \[MS_W = \frac{1}{J(\bar{n}-1)}\sum_{j=1}^J\sum_{i=1}^{n_j}(Y_{ij}-\bar{Y}_{\cdot j})^2,\text{ where}\quad \bar{n} = \frac{1}{J}\sum_{j=1}^J n_j,\]
and the mean square between groups $MS_B$ by \[MS_B = \frac{1}{J-1}\sum_{j=1}^J\sum_{i=1}^{n_j}(\bar{Y}_{\cdot j}-\bar{Y}_{\cdot\cdot})^2,\text{ where}\quad \bar{Y}_{\cdot\cdot}=\frac{\sum_{j=1}^J \frac{n_j}{\sigma^2}\bar{Y}_{\cdot j}}{\sum_{j=1}^J \frac{n_j}{\sigma^2}}.\]
Then, unbiased estimators for $\sigma^2$ and $\tau^2$ are given by $\hat{\sigma}^{2\text{ANOVA}}=MS_W$ and $\hat{\tau}^{2\text{ANOVA}} = \frac{MS_B-MS_W}{\bar{n}}$.

The same authors prove that
\[p(\tau|\sigma^2,\Y) \propto p(\tau)V_\mu^{1/2}\prod_{j=1}^J (\bar\sigma_j^2+\tau^2)^{-1/2}\exp\llavs{-\frac{(\bar{Y}_{\cdot j}-\hat{\mu})}{2(\bar\sigma_j^2+\tau^2)}}.\]
Note that everything multiplying $p(\tau)$ approaches a nonzero constant limit as $\tau$ tends to zero. Thus, the behavior of the posterior density near $\tau=0$ is determined by the prior density. The usual noninformative function $p(\tau)\propto \frac{1}{\tau}\indicator{(0,\infty)}{(\tau)}$ is not integrable for any small interval including $\tau = 0$  and yields a nonintegrable posterior density. Meanwhile, the uniform prior distribution $p(\tau)\propto \indicator{(0,\infty)}{(\tau)}$ yields a proper posterior density.

\begin{figure}[ht]
	\centering
	\includegraphics[width=0.65\textwidth]{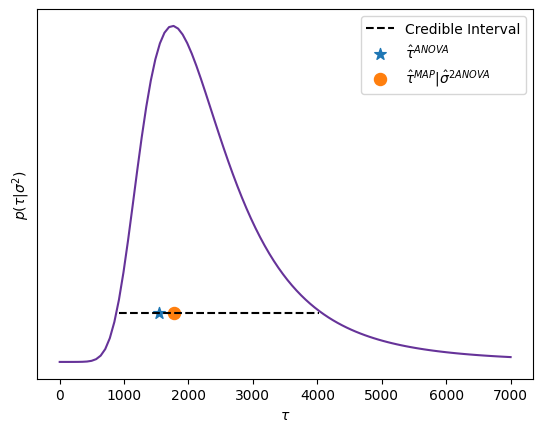}
	\caption{Posterior distribution of $\tau$ conditional on $\sigma^2$. We have set $\sigma^2$ to its empirical estimator. We show the empirical estimator for $\tau$ based on the ANOVA and the mode of the distribution. Both estimators are presented on an interval of approximately 0.95 posterior probability, we calculate this interval based on the asymptotic behavior of the posterior distribution.}
	\label{fig:TauDistributionEmpirical}
\end{figure}

In \Cref{fig:TauDistributionEmpirical} we present the posterior conditional distribution of $\tau$ conditional on the empirical estimate of $\sigma^2$, that is $p(\tau|\hat{\sigma}^{2\text{ANOVA}})$. We also present the mode of this distribution, and an approximate interval of 0.95 probability. Under conditions of regularity a $(1-\alpha)\times 100\%$ probability interval is given by
\[\llavs{\tau:\frac{p(\tau|\sigma^2,\Y)}{p(\hat{\tau}^{\text{MAP}}|\sigma^2,\Y)}\geq\exp\llavs{-\frac{q_{\chi^2_1}^{1-\alpha}}{2}}},\]
where $q_{\chi^2_1}^{1-\alpha}$ denotes the quantile of probability $1-\alpha$ of a $\chi^2$ distribution with 1 degree of freedom.

\begin{figure}[ht]
	\centering
	\includegraphics[width=0.65\textwidth]{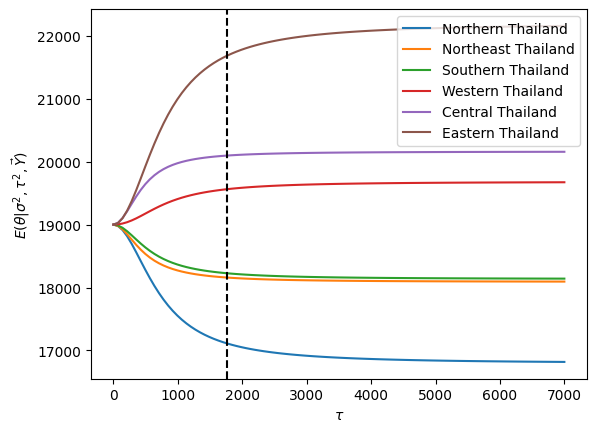}
	\caption{Conditional posterior mean for $\theta_j$, $\E(\theta_j|\tau,\sigma^2,\Y)$ ($j=1,\ldots,J$) as function of $\tau$, setting $\sigma^2$ to $\hat{\sigma}^{2\text{ANOVA}}$. The vertical dashed line is plot on $\hat{\tau}^{\text{MAP}}$.}
	\label{fig:ThetaEmpirical}
\end{figure}

In \Cref{fig:ThetaEmpirical} we present $\E(\theta_j|\tau,\sigma^2,\Y)$, calculated in \Cref{eq:ETheta}, as a function of $\tau$ and setting the value of $\sigma^2$ to $\hat{\sigma}^{2\text{ANOVA}}$. We plot a vertical dashed line on the value $\hat{\tau}^{\text{MAP}}$. We can observe how this estimated value lies between the extreme cases of the complete pooling and no pooling models.

Consider now the model presented in \Cref{subsec:HierVaryingSigma}, and remember that
\[\sigma_j^2|\bm{\theta},\nu,\rho^2,\Y\sim\textsf{Inverse-}\chi^2(\nu_j,\hat{\sigma}_j^2),\]
where
\[\nu_j=\nu+n_j,\quad \hat{\sigma}_j^2=\frac{\nu\rho^2+n_jv_j}{\nu+n_j},\text{and }v_j=\frac{1}{n_j}\sum_{i=1}^{n_j}(Y_{ij}-\theta_j)^2.\]
Then,
\begin{align*}
\E(\sigma_j^2|\bm{\theta},\nu,\rho^2,\Y) & = \frac{\nu_j}{\nu_j-2}\hat{\sigma}_j^2 \\
& = \frac{\nu\rho^2+n_jv_j}{\nu+n_j-2},
\end{align*}
and
\begin{align*}
\V(\sigma_j^2|\bm{\theta},\nu,\rho^2,\Y) &= \frac{2\nu_j^2}{(\nu_j-2)^2(\nu_j-4)}\hat{\sigma}_j^4 \\
& = \frac{2(\nu\rho^2+n_jv_j)^2}{(\nu+n_j-2)^2(\nu+n_j-4)},
\end{align*}
from this expressions is easy to see that $\E(\sigma_j^2|\bm{\theta},\nu,\rho^2,\Y)\xrightarrow[\nu\to \infty]{} \rho^2$ and $\V(\sigma_j^2|\bm{\theta},\nu,\rho^2,\Y)\xrightarrow[\nu\to \infty]{} 0$. On the other hand, if $\nu\to 0$, then $\sigma_j^2|\bm{\theta},\nu,\rho^2,\Y\sim\textsf{Inverse-}\chi^2(n_j,v_j)$, corresponding with the no pooling inference.

\begin{figure}[ht]
	\centering
	\includegraphics[width=0.65\textwidth]{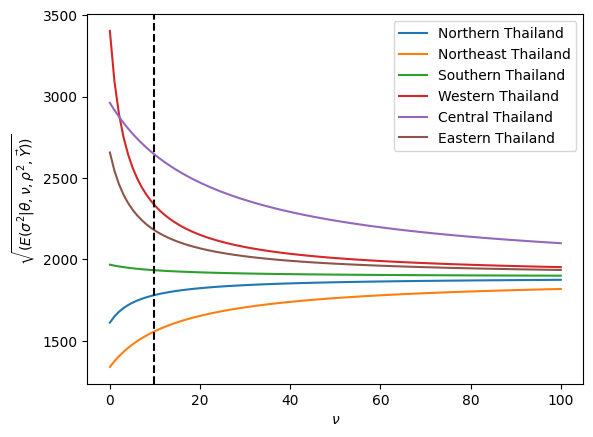}
	\caption{Square root of the conditional posterior mean for $\sigma^2_j$, $\sqrt{\E(\sigma_j^2|\bm{\theta},\nu,\rho^2,\Y)}$ ($j=1,\ldots,J$) as function of $\nu$, setting $\rho^2$ to $\hat{\rho}^2$ and substituting $v_j$ for the observed variance in the region. The vertical dashed line is plot on $\hat{\nu}$.}
	\label{fig:SigmaEmpirical}
\end{figure}

In \Cref{fig:SigmaEmpirical} we present $\sqrt{\E(\sigma_j^2|\bm{\theta},\nu,\rho^2,\Y)}$, as a function of $\nu$ and setting the value of $\rho^2$ to the estimator $\hat{\rho}^2$ calculated with \Cref{eq:HatRho}, and substituting $v_j$ for the observed variance. We plot a vertical dashed line on the value $\hat{\nu}$ (see \Cref{eq:HatNu}). Once again, we can observe how this estimated value lies between the extreme cases of complete pooling and no pooling models.

\begin{figure}[ht]
	\begin{center}
        \scalebox{.9}{
		\begin{tikzpicture}[>=stealth', decoration={markings, mark=at position 1 with {\arrow[scale=2]{>}}}]
			\draw[postaction={decorate},->] (-0,0) -- (11,0); 
			\draw[postaction={decorate},->] (0,-0) -- (0,6); 
			
			\draw (0,0) node[point, above right] {Complete pooling};
			\draw (0,0) node[below right] {1410.13};
			
			\draw (5.5,0) node[point, below] {Hierarchical model with common $\sigma^2$};
			\draw (5.5,-0.5) node[below] {1386.44};
			
			\draw (5.5,2) node[point, right] {$h=1$};
			\draw (5.5,2) node[left] {1380.31};
			
		
			\draw (5.5,3) node[point, right] {$h=2$};
			\draw (5.5,3) node[left] {1380.15};
			
			\draw (5.5,4) node[point, right] {$h=3$};
			\draw (5.5,4) node[left] {1380.30};
			
			\draw (9.5,6) node[point, below] {No pooling};
			\draw (9.5,5.5) node[below] {1382.01};

			\draw[postaction={decorate},->] (2.5,5) -- (4,2.5);
			\draw (2.5,5.5) node[above] {$\nu$ fixed at $\hat{\nu}$};
			\draw (2.5,5.5) node[below] {\textbf{1378.58}};

			\draw[postaction={decorate},->] (9,2.5) -- (7,2.5);
			\draw (9.5,2.5) node[above] {$\nu\sim\textsf{Exponential}(1/\hat{\nu})$};
			\draw (9,2.5) node[below] {1379.21};
			
			\draw [dashed] (5.5,2.75) ellipse (2.00cm and 2.75cm);
			
			\draw[postaction={decorate},->] (5.5,1.5) -- node[right] {$h\to 0$} (5.5,0.5);

			\draw[postaction={decorate},->] (-0.5,3) -- node[above, rotate=90] {$\nu\to \infty$} (-0.5,0.5);
			\draw[postaction={decorate},->] (-0.5,3.5) -- node[above, rotate=90] {$\nu\to 0$} (-0.5,6);
			\draw[postaction={decorate},->] (-1.5,3) -- node[above, rotate=90] {same $\sigma_j^2$} (-1.5,0.5);
			\draw[postaction={decorate},->] (-1.5,3.5) -- node[above, rotate=90] {different $\sigma_j^2$} (-1.5,6);
			\draw[postaction={decorate},->] (-2.5,1) -- node[above, rotate=90] {more variance in $\sigma_j^2$} (-2.5,5);
			\draw[postaction={decorate},->] (-3.5,1) -- node[above, rotate=90] {less bias in $\sigma_j^2$} (-3.5,5);

			\draw[postaction={decorate},->] (3,-1) -- node[above] {$\tau\to 0$} (0.5,-1);
			\draw[postaction={decorate},->] (8,-1) -- node[above] {$\tau\to \infty$} (10.5,-1);
			\draw[postaction={decorate},->] (3,-2) -- node[above] {same $\theta_j$} (0.5,-2);
			\draw[postaction={decorate},->] (8,-2) -- node[above] {different $\theta_j$} (10.5,-2);
			\draw[postaction={decorate},->] (3.5,-3) -- node[above] {more variance in $\theta_j$} (7.5,-3);
			\draw[postaction={decorate},->] (3.5,-4) -- node[above] {less bias in $\theta_j$} (7.5,-4);
		\end{tikzpicture}
        }
		\caption{Relation between models and the influence of the different parameters in the qualitative characteristics of the models. Next to each model we present its corresponding WAIC value for the monthly income, presented in \Cref{tab:WaicAllModels}.}
		\label{fig:RelationBetweenModels}
	\end{center}
\end{figure}
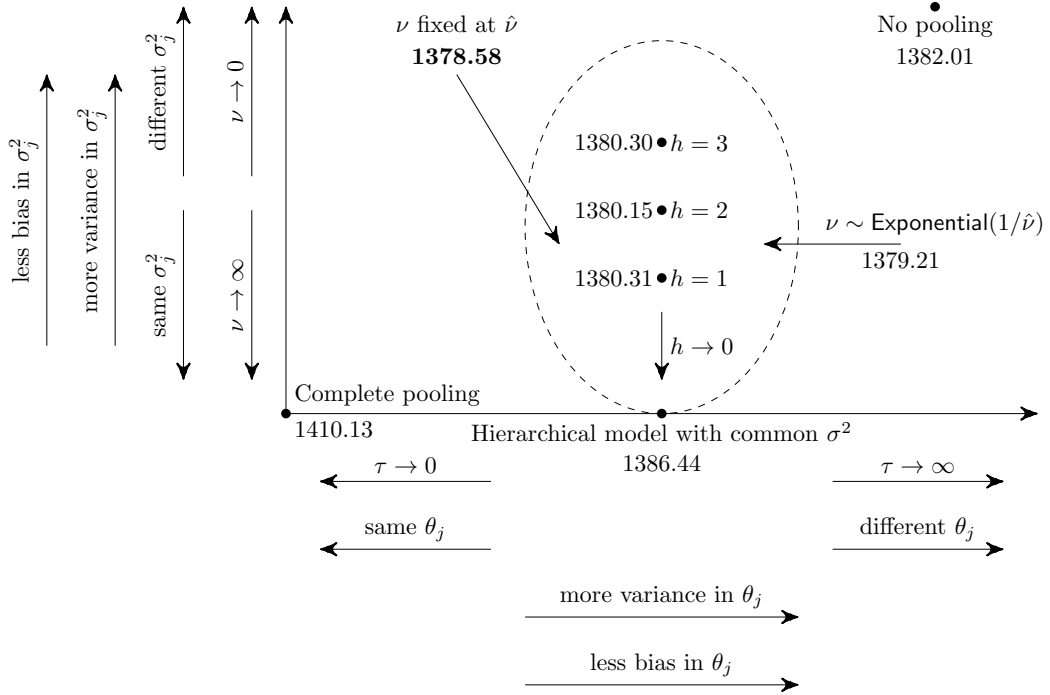

In \Cref{fig:RelationBetweenModels} we present a diagram representing the different models considered in \Cref{sec:HierRegion} and their relation with the parameters. On the bottom left of the diagram we present the complete pooling model, which assumes the same within variance for all the regions and the same mean. While on the other extreme of the diagram we present the no pooling model, where separate independent models are adjusted for each region. The hierarchical models are between these two extreme cases. When we consider $p(\nu)\propto \nu^{-h}\indicator{(0,\infty)}{(\nu)}$, we observed trough simulations that large values of $h$ tend to generate within-region variances similar to the no pooling model, while a common within-region variance is obtained when $h\to 0$. Somewhere around these models are our other two approaches, set $\nu$ to a fixed estimated value $\hat{\nu}$ or assign a prior exponential distribution. Next to each model we present its corresponding WAIC value for the monthly income variable, showing in bold the model with the lowest WAIC (see \Cref{tab:WaicAllModels}).

\subsection{Proofs}

\subsubsection{Conditional posterior distribution of \texorpdfstring{$\rho^2$}{rho2}}
\label{subsubsec:PosteriorRho2}

To calculate the conditional posterior distribution of $\rho^2$, note that
\begin{align*}
p(\rho^2|\bm{\sigma}^2,\nu) & \propto p(\rho^2|\nu)p(\bm{\sigma}^2|\nu,\rho^2) \\
& \propto p(\rho^2)\prod_{j=1}^J (\rho^2)^{\nu/2}\exp\llavs{-\frac{\nu}{2\sigma_j^2}\rho^2} \\
& = p(\rho^2) (\rho^2)^{\frac{J\nu}{2}}\exp\llavs{-\parent{\frac{\nu}{2}\sum_{j=1}^J \frac{1}{\sigma_j^2}}\rho^2}.
\end{align*}
Setting $p(\rho^2)=\frac{1}{\rho^2}\indicator{(0,\infty)}{(\rho^2)}$ yields
\[p(\rho^2|\bm{\sigma}^2,\nu)\propto (\rho^2)^{\frac{J\nu}{2}-1}\exp\llavs{-\parent{\frac{\nu}{2}\sum_{j=1}^J \frac{1}{\sigma_j^2}}\rho^2}\indicator{(0,\infty)}{(\rho^2)}.\]
It is immediate from the previous expression that
\[\rho^2|\bm{\sigma}^2,\nu\sim\textsf{Gamma}\parent{\frac{J\nu}{2},\frac{J\nu}{2\hat{\rho}^2}},\]
where \[\hat{\rho}^2 = \frac{J}{\sum_{j=1}^J \frac{1}{\sigma_j^2}}.\]
Note that $\E(\rho^2|\bm{\sigma}^2,\nu)=\hat{\rho}^2$ is the harmonic mean of the within-groups variances. Thus $\rho^2$ models the ``common'' within-variance.

\subsubsection{Estimating \texorpdfstring{$\rho^2$}{rho2} and \texorpdfstring{$\nu$}{nu}}
\label{subsubsec:EstimateRhoNu}

Because $\sigma_j^2\sim\textsf{Inverse-}\chi^2(\nu,\rho^2)$, then \[\E(\sigma_j^2|\nu,\rho^2)=\frac{\nu}{\nu-2}\rho^2\] and \[\V(\sigma_j^2|\nu,\rho^2)=\frac{2\nu^2}{(\nu-2)^2(\nu-4)}\rho^4.\] Let be $E_{s^2}$ the average of the observed sample within-group variances $s_1^2,\ldots,s_J^2$ and $V_{s^2}$ their variance. Then, using the method of moments we have
\[E_{s^2}=\frac{\hat{\nu}}{\hat{\nu}-2}\hat{\rho}^2\Rightarrow\hat{\rho}^2=\frac{\hat{\nu}-2}{\hat{\nu}}E_{s^2},\]
and
\begin{align*}
V_{s^2} &= \frac{2\cancel{\hat{\nu}^2}}{\cancel{(\hat{\nu}-2)^2}(\hat{\nu}-4)}\frac{\cancel{(\hat{\nu}-2)^2}}{\cancel{\hat{\nu}^2}}(E_{s^2})^2 \\
\Rightarrow \hat{\nu} &= \frac{2(E_{s^2})^2}{V_{s^2}}+4,
\end{align*}
thus
\begin{align*}
\hat{\rho}^2 &= \parent{1-\frac{2V_{s^2}}{2(E_{s^2})^2+4V_{s^2}}}E_{s^2} \\
& = \parent{\frac{2(E_{s^2})^2+2V_{s^2}}{2(E_{s^2})^2+4V_{s^2}}}E_{s^2}.
\end{align*}

\section{Supplementary Figures}
\label{sec:AdditionalFigures}

\subsection{Income per region and education level}

\begin{figure}[ht]
	\centering
	\includegraphics[width=0.85\textwidth]{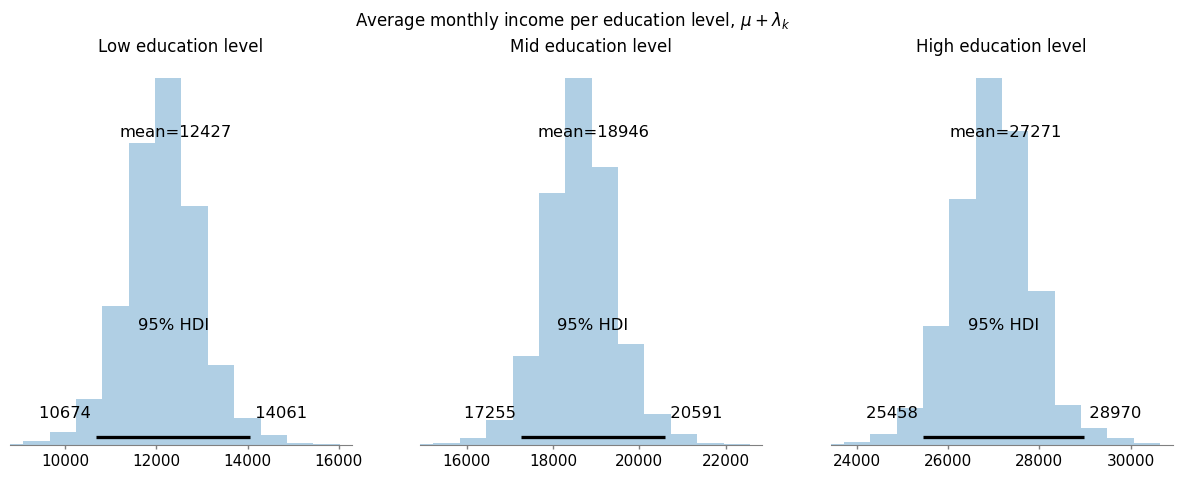}
	\caption{Average income per education level.}
	\label{fig:IncomePerEducationLevel}
\end{figure}

\begin{figure}[ht]
	\centering
	\includegraphics[width=0.85\textwidth]{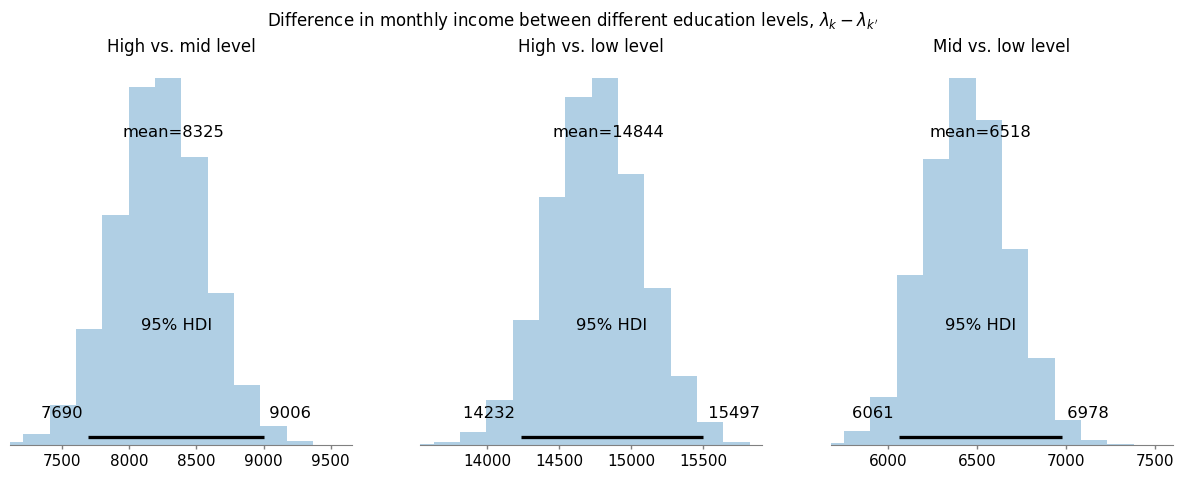}
	\caption{Difference between average income per education level.}
	\label{fig:IncomePerEducationLevelDifference}
\end{figure}

Remember that the expected monthly income for the region $j$ and the education level $k$ is given by $\theta_j+\lambda_k$, where $\theta_j|\mu,\tau^2\sim\Normal(\mu,\tau^2)$. Then, fixing the education level, and taking the average over all the regions, i.e. computing the expected value over $\theta_j$, we obtain the expected value of the monthly income for the education level $k$, given by $\mu+\lambda_k$. We present in \Cref{fig:IncomePerEducationLevel} the expected monthly income for each education level. Similarly, \Cref{fig:IncomePerEducationLevelDifference} shows the difference of income between the education levels. We can conclude from these Figures that the education level has an important impact on the income of households, when averaging over all the regions, a higher education level is related with a higher income.

\begin{figure}[ht]
	\centering
	\includegraphics[width=0.9\textwidth]{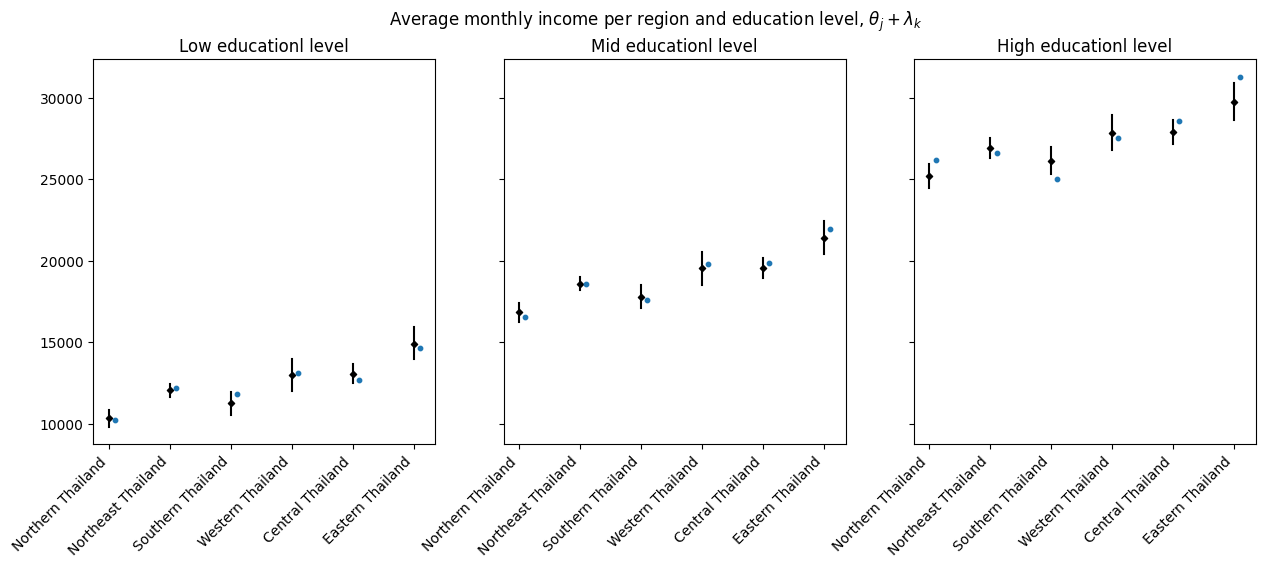}
	\caption{Average monthly income per region and education level.}
	\label{fig:MeanIncomePerRegionAndEducationLevel}
\end{figure}

\begin{figure}[ht]
	\centering
	\includegraphics[width=0.9\textwidth]{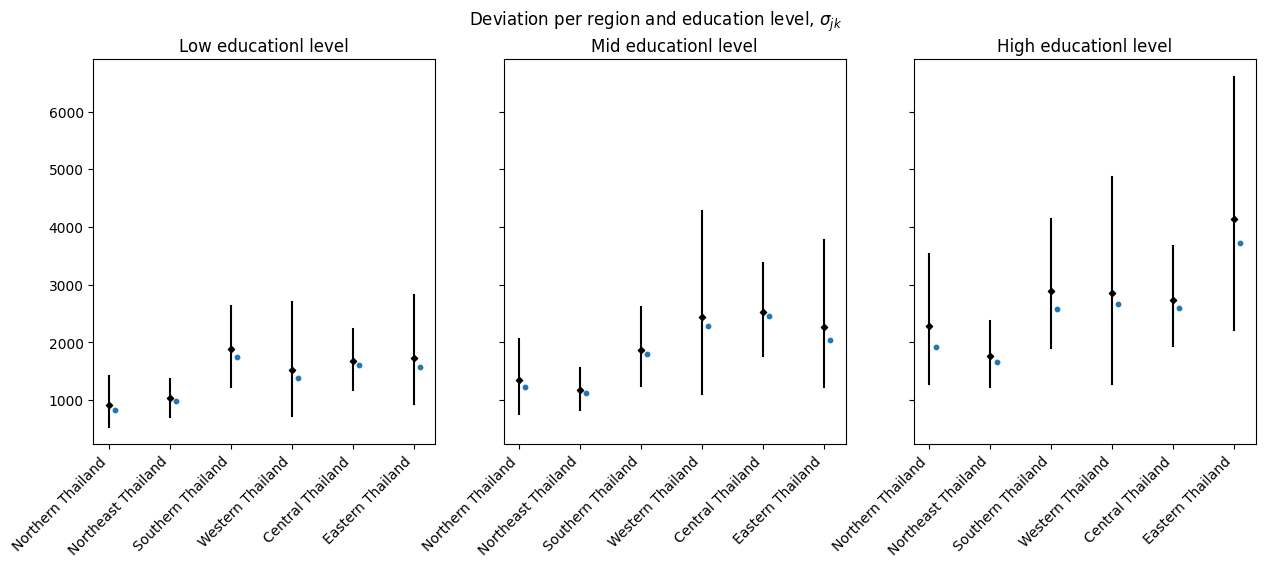}
	\caption{Deviation within-region of the monthly income per region and education level.}
	\label{fig:DeviationIncomePerRegionAndEducationLevel}
\end{figure}

\begin{figure}[ht]
	\centering
	\includegraphics[width=0.9\textwidth]{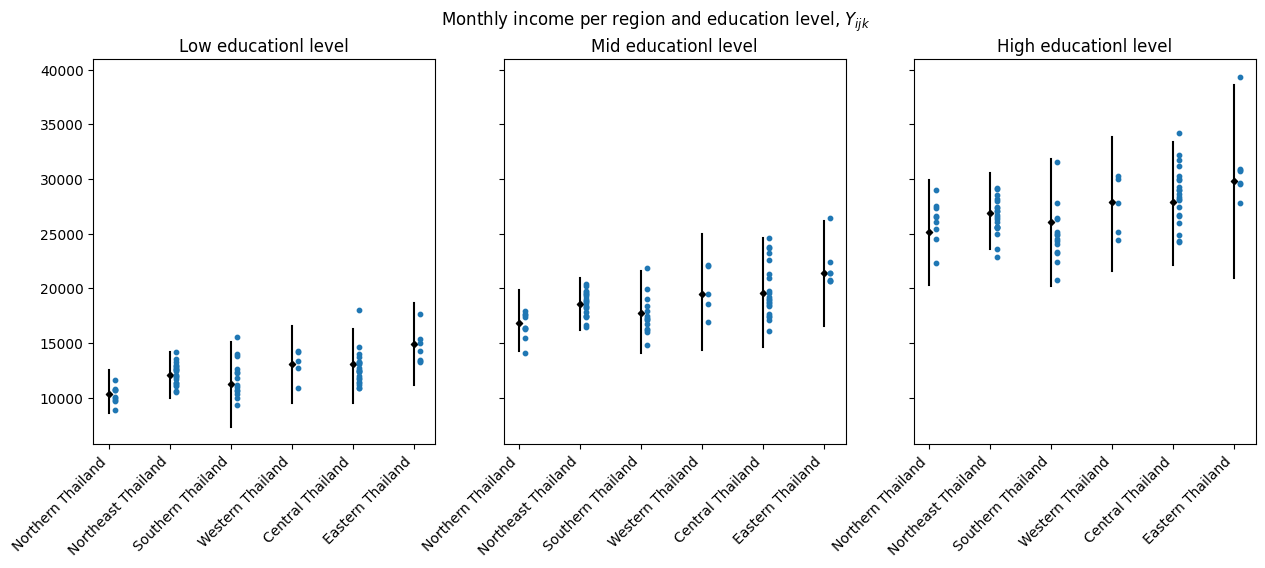}
	\caption{Province average monthly income per region and education level.}
	\label{fig:YIncomePerRegionAndEducationLevel}
\end{figure}

We present in \Cref{fig:MeanIncomePerRegionAndEducationLevel,fig:DeviationIncomePerRegionAndEducationLevel,fig:YIncomePerRegionAndEducationLevel} our ubiquitous graphs showing the regional average, the regional deviation and the province average of the monthly income. This time, however, we present these quantities for each education level. We observe that the education level has an important impact on income regardless of the region.

\end{appendix}

\end{document}